\documentclass[aps,prl,reprint,twocolumn]{revtex4-1}
\usepackage{graphicx}
\usepackage{amsmath}
\usepackage{amssymb}
\usepackage[colorlinks,allcolors=blue]{hyperref}
\usepackage{hypcap}

\newcommand{\pref}[2]{\hyperref[#1]{\ref{#1}#2}}
\newcommand{\eqpref}[1]{\hyperref[#1]{(\ref{#1})}}

\begin{document}
\title{Correlated dynamics in a synthetic lattice of momentum states}
\author{Fangzhao~Alex~An}
\thanks{These authors contributed equally to this work}
\author{Eric~J.~Meier}
\thanks{These authors contributed equally to this work}
\author{Jackson~Ang'ong'a}
\author{Bryce~Gadway}
\email{bgadway@illinois.edu}
\affiliation{Department of Physics, University of Illinois at Urbana-Champaign, Urbana, IL 61801-3080, USA}
\date{\today}
\begin{abstract}
We study the influence of atomic interactions on quantum simulations in momentum-space lattices (MSLs), where driven transitions between discrete momentum states mimic transport between sites of a synthetic lattice. Low energy atomic collisions, which are short ranged in real space, relate to nearly infinite-ranged interactions in momentum space. However, the added exchange energy between atoms in distinguishable momentum states leads to an effectively attractive, finite-ranged interaction in momentum space.
In this work, we observe the onset of self-trapping driven by such interactions in a momentum-space double well, paving the way for more complex many-body studies in tailored MSLs.
%In a momentum-space double well, we observe the onset of self-trapping driven by such interactions, paving the way for more complex many-body studies in tailored MSLs.
We consider the types of phenomena that may result from these interactions, including the formation of chiral solitons in topological zigzag lattices.
\end{abstract}
\maketitle

Quantum simulation with ultracold atoms~\cite{Bloch-RMP08,AtomsRev-NatPhys-2012} has been a powerful tool in the study of many-body physics and nonequilibrium dynamics. There has been recent interest in extending quantum simulations from real-space potentials to synthetic lattice systems composed of discrete internal~\cite{Boada-Synth,Celi-ArtificialDim} or external~\cite{NateGold-TrapShake} states. These synthetic dimensions enable many unique capabilities for quantum simulation, including the ability to engineer nontrivial topology~\cite{Celi-ArtificialDim,Wall-SpinOrb} and higher dimensions~\cite{Boada-Synth}.

Our recent development of momentum-space lattices (MSLs), based on the use of discrete momentum states as effective sites, has introduced a fully synthetic approach to simulating lattice dynamics~\cite{Gadway-KSPACE,Meier-AtomOptics,Meier-SSH,Alex-2Dchiral,Alex-Annealed}. As compared to partially synthetic approaches~\cite{Fallani-chiral-2015,Stuhl-Edge-2015}, fully synthetic lattices offer the possibility of studying coherent internal-state dynamics that are decoupled from any motional entropy~\cite{yan2013:dipole-dipole}. Moreover, synthetic lattices offer a microscopic control over all system parameters, analogous to that found in photonic simulators~\cite{SzameitReview-2010,PhotRev-NatPhys-2012} but with a more natural path to exploring the influence of interactions.

Still, fully synthetic systems face some challenges in exploring nontrivial atomic interactions. Synthetic systems based purely on internal states suffer from limited state spaces, sensitivity to external noise for generic, field-sensitive states~\cite{Sugawa-NonAb}, and loss due to two- and three-body collisions~\cite{Soding-relaxation,Weiner-Collisions}. Furthermore, for atoms featuring spin-independent scattering lengths (as in the alkaline earths~\cite{Pagano-AlkalineEarth} and roughly for $^{87}$Rb~\cite{Sugawa-NonAb}), interactions are independent of the internal state distribution and are thus decoupled from fully synthetic internal-state dynamics. Similarly, the nearly all-to-all momentum-space interactions that result from short-ranged collisions in real space should naively be decoupled from MSL dynamics.

Here, we show that finite-ranged interactions appear in MSL studies, resulting from a combination of momentum-independent scattering and the exchange energy of bosonic atoms in distinguishable momentum states. We experimentally probe correlated dynamics driven by these interactions, observing the onset of nonlinear self-trapping in a double well system. Considering the influence of interactions on dynamics in tailored MSLs, we show through simulations that stable chiral solitons emerge in the presence of a magnetic flux.

MSLs provide a bottom-up approach to engineering designer Hamiltonians with field-driven transitions. This technique utilizes the coherent coupling of multiple atomic momentum states via two-photon Bragg transitions to synthesize an effective lattice of coupled modes in momentum space. The resonance frequency associated with each Bragg transition is generally unique. For free non-interacting particles, this stems from the quadratic dispersion $E_p^0 = p^2/2m$, with momentum $p$ and atomic mass $m$. Considering atoms initially at rest and driven by a single pair of counter-propagating lasers with wavelength $\lambda$ and wavevector $k = 2\pi/\lambda$, a discrete set of momentum states $p_n = 2n\hbar k$ may be coupled, having energies $4 n^2 E_R$, with $E_R = \hbar^2 k^2 / 2m$ being the photon recoil energy and with $n$ being the integer state index. Individual addressing of the unique Bragg transitions allows us to engineer, with full local and temporal parameter control, the single-particle tight-binding model
\begin{equation}
H^\text{sp} \approx -\sum_{n,\alpha} t_{n,\alpha}(e^{i \varphi_{n,\alpha}} \hat{c}^\dag_{n+\alpha} \hat{c}_n + \mathrm{h.c.}) + \sum_n \varepsilon_n \hat{c}^\dag_n \hat{c}_n \ ,
\label{EQ:e00}
\end{equation}
where $\hat{c}_n$ ($\hat{c}^\dagger_n$) is the annihilation (creation) operator for the state with momentum $p_n$. Here, tunneling of order $\alpha$
(e.g. $\alpha = 1$ and $2$ for nearest- and next-nearest-neighbor tunnelings, respectively)
%(e.g. $\alpha = 1$ for nearest-neighbor tunneling, $\alpha = 2$ for next-nearest-neighbor)
is controlled through the amplitudes and phases of individual frequency components used to drive Bragg transitions of order $\alpha$~\cite{Kozuma-Bragg}. Similarly, an effective potential landscape of site energies $\varepsilon_n$ is controlled by small detunings from Bragg resonance.

While MSLs have seen success in engineering diverse single-particle Hamiltonians~\cite{Meier-AtomOptics,Meier-SSH,Alex-2Dchiral,Alex-Annealed}, the prospects for studying interactions and correlated dynamics have not yet been examined. In real-space atomic systems, correlated physics has largely been driven by two-body contact interactions~\cite{Greiner-SF-MI-2002,Greiner-AFM}, which are nearly infinite ranged in momentum space (i.e. with an energy independent of relative momentum for two colliding, distinguishable atoms). In one dimension, where collisions are mode-preserving, the resulting all-to-all interactions in MSLs would appear incapable of driving correlated behavior.

\begin{figure}[t!]
\includegraphics[width=\columnwidth]{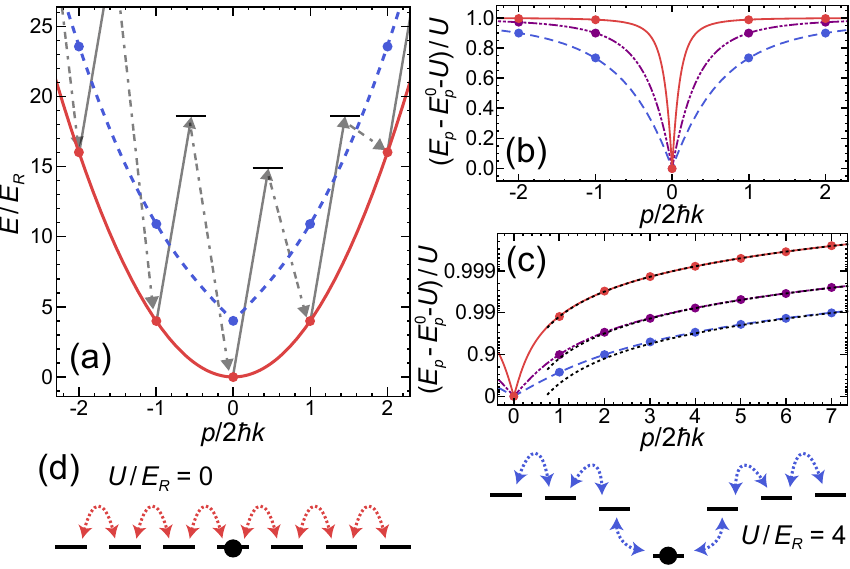}
\caption{\label{FIG:disp}
\textbf{Interaction shifts of Bragg tunneling resonances.}
\textbf{(a)}~Energy dispersion of non-interacting massive atoms $E_p^0$ (red solid line) in units of the recoil energy $E_R$, and the Bogoliubov dispersion $E_p$ of a homogeneous gas with weak repulsive interactions and a mean-field energy $U = 4 E_R$ (blue dashed line).
\textbf{(b)}~The effective momentum-space lattice site energies (with a common shift of $U$ removed and renormalized to $U$) experienced by weakly-coupled excitations of a macroscopically populated $p=0$ condensate, shown for the cases $U / E_R = \ $0.1, 1, and 4 (solid red, dash-dotted purple, and dashed blue lines, respectively).
\textbf{(c)}~Curves as in  (b), but shown on a semi-log scale and for a larger range of momenta, compared to the form $U - U^2/2 E_p^0$ (dotted black lines) relevant in the free-particle limit ($E_p^0 \gg 2U$).
\textbf{(d)}~Depiction of effective site energies shifted by interactions with a $p=0$ condensate for $U/E_R = 0$ and $U/E_R = 4$.
}
\end{figure}

This description is incomplete, however, as the quantum statistics of identical bosonic atoms in MSLs create a structured interaction profile in momentum space. Identical bosons in distinguishable motional states experience an added exchange interaction owing to the symmetry properties of the two-boson wave function~\cite{Ozeri-Bog-RMP,exch,Kaufman-Entang,Larson-MultibandBosons}.
Given repulsive real-space interactions, this additional long-ranged momentum-space repulsion between atoms in different momentum states can alternatively be viewed as an effective attraction between atoms in the same momentum state, an interaction which promotes Bose--Einstein condensation in weakly interacting gases~\cite{Smith-Interactions-BEC}.
This combination of momentum-independent collisions and quantum statistics results in a number of hallmark features of bosonic quantum fluids, such as the Bogoliubov quasiparticle dispersion and distinct transport properties of heat and sound~\cite{SecondSound}.

To gain some intuition for the form of interactions in MSLs, we consider the Bogoliubov spectrum~\cite{Stenger-Bragg,Ozeri-Bog-RMP,Hadzibabic-Bragg-Strong} for the limiting case of all population initially residing at a single site.
%Consideration of the Bogoliubov spectrum~\cite{Stenger-Bragg,Ozeri-Bog-RMP,Hadzibabic-Bragg-Strong} helps us gain some intuition for the form of interactions in MSLs, at least in the limiting case that all population initially resides at a single site.
We assume a uniform number density $\rho_N$ relating to a homogeneous mean-field energy $U = g \rho_N$, for interaction parameter $g = 4\pi \hbar^2 a/m$ and $s$-wave scattering length $a$.
While repulsive interactions raise the energy of $p=0$ condensate atoms by this mean-field energy $U$, high-momentum excitations ($E_p^0 \gg 2U$) experience an interaction energy shift of roughly $2U$ due to both direct and exchange interactions with the $p=0$ condensate. For a general momentum $p$, the Bogoliubov quasiparticle excitations have an energy $E_p = U + \sqrt{E_p^0 (E_p^0 + 2 U)}$. Figure~\ref{FIG:disp} depicts this modified energy dispersion, along with the form of the effective interaction-dependent shifts to the MSL site energies, which relate to the difference in energy between the final state and the initial $p=0$ state.
%For $\mu \gtrsim E_R$, the momentum-space interaction potential has significant off-site contributions, the effective range of which increases with larger ratios $\mu / E_R$. There is, however, a natural limitation on the compatibility of long-ranged interactions with the scheme for engineering MSLs. This method breaks down when unique spectral addressing of the individual Bragg transitions is lost, occurring when multiple momentum orders populate the linear phonon branch (occurring roughly when $\mu$ exceeds $8 E_R$, the bare energy spacing of the Bragg resonances).

\begin{figure*}[t!]
\includegraphics[width=\textwidth]{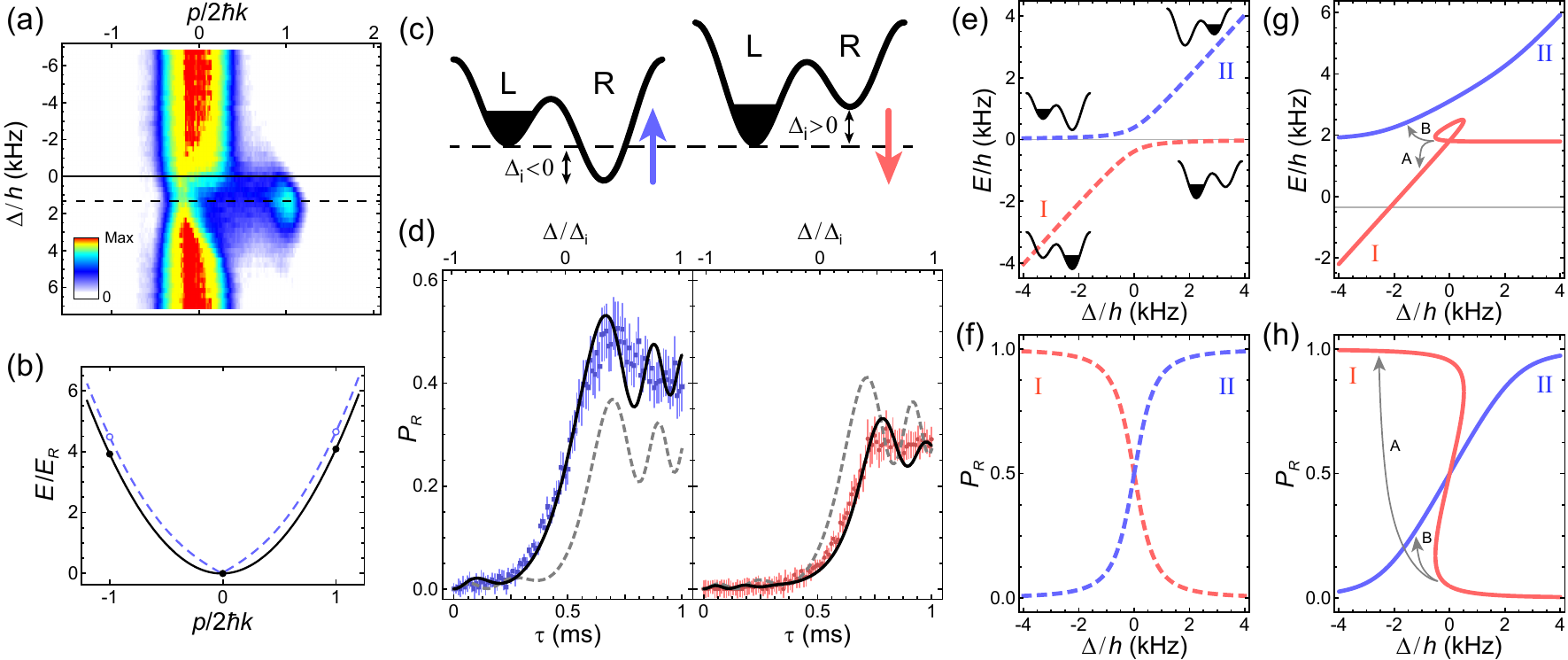}
\caption{\label{FIG:expt}
\textbf{Interaction effects in a momentum-space double well.}
\textbf{(a)}~Bragg spectroscopy of the $0 \rightarrow 1$ transition showing an interaction-driven shift (dashed line) of $1.31(3)$~kHz from single particle resonance (solid line).
The momentum distributions, relating to the integrated optical density after 18~ms time of flight, reveal the fraction of population transferred to the state $p = 2\hbar k$ by a 400~$\mu$s-long Bragg pulse as a function of detuning from single-particle resonance, relating to an intersite energy bias $\Delta$.
\textbf{(b)}~Measured shifts for both $0\rightarrow\pm 1$ transitions overlaid onto the Bogoliubov dispersion (dashed blue) with single-particle dispersion for comparison (solid black). The Bogoliubov dispersion has been shifted by $-0.018\hbar k$ to account for initial condensate momentum.
\textbf{(c)}~Experimental protocol for double well sweeps. Population begins in left well ($L$) and transfers to the right well ($R$) as the imbalance $\Delta$ (detuning from single-particle resonance) is swept across 0 in the positive direction (negative to positive, blue arrow at left) and in the negative direction (positive to negative, red arrow at right). Dashed line indicates zero energy.
\textbf{(d)}~Population in the right well $P_R$ vs. time $\tau$ during the 1~ms-long linear bias sweeps (lower horizontal axis), or equivalently the instantaneous ratio of the bias to the initial bias, $\Delta / \Delta_i$, throughout the sweep (upper horizontal axis). The positive and negative sweeps are shown separately on the left (blue squares) and the right (red dots), respectively. Dashed gray curves represent single-particle predictions, and solid black curves represent predictions taking into account an inhomogeneous density distribution with an average mean-field energy of $U/\hbar \approx 2\pi\times1.81$~kHz~\cite{Inter-SuppMat}.
\textbf{(e)}~Adiabatic energy levels (I and II) of the double well without interactions as a function of $\Delta$. Cartoon insets depict the population distributions for large $|\Delta / t|$.
\textbf{(f)}~Population projection of the adiabatic levels in (\textbf{e}) onto the right well as a function of $\Delta$ without interactions.
\textbf{(g,h)}~Adiabatic energy levels and right-well population projections as in (\textbf{e,f}), but with an added homogeneous mean-field energy of $U/\hbar \approx 2\pi\times1.81$~kHz. Arrows A and B on the negative sweep denote forced tunneling pathways as the population transfer overshoots due to self-trapping.
Error bars in (\textbf{b}) and (\textbf{d}) denote one standard error of the mean.
}
\end{figure*}

In our system of Bose-condensed $^{87}$Rb atoms~\cite{Meier-AtomOptics}, we can directly probe the interaction energy shifts of the momentum states through Bragg spectroscopy~\cite{Stenger-Bragg,Ozeri-Bog-RMP,Hadzibabic-Bragg-Strong}. For our laser wavelength of $1064$~nm, first-order Bragg resonances are expected at frequency detunings of $\pm 4 E_R/\hbar \approx \pm 2\pi \times 8.1$~kHz for non-interacting atoms.
As displayed in Figs.~\pref{FIG:expt}{(a,b)}, we measure an average shift of these first-order Bragg resonances by $2\pi \times 1.14(5)$~kHz.
%for a tunneling rate $t/\hbar \approx 2\pi\times 390$~Hz.
The shifts to the $+1$ and $-1$ transitions are slightly different due to nonzero initial momentum of the condensate, and we have shifted the Bogoliubov dispersion in Fig.~\pref{FIG:expt}{(b)} to account for this.
%For a uniform gas, this shift would relate to a homogeneous mean-field energy $U/\hbar \approx 2\pi\times 1.81$~kHz.
%In our harmonic trap, $U$ relates instead to the average mean-field shift in the trapped gas, with local mean-field energies ranging from zero to $U_0/\hbar \approx 2\pi\times 3.17$~kHz in the central region with peak density~\cite{Stenger-Bragg}.

As a first experimental study, we explore the influence of momentum-space interactions on population dynamics in a coupled double well. We initialize all of the population in the left well ($p = 0$ state), with a large initial energy bias $\Delta_i$ between the left (L) and right (R) wells that inhibits tunneling. The bias $|\Delta_i|/\hbar = 2\pi \times 8$~kHz is chosen to greatly exceed the tunneling energy $t/\hbar \approx 2\pi \times 390$~Hz. As depicted in Fig.~\pref{FIG:expt}{(c)}, the bias is then linearly swept through zero (single particle resonance) to a final value $\Delta_f = -\Delta_i$ over a duration of 1~ms. We consider both the case of a positive sweep (from $\Delta_i < 0$ to $\Delta_f > 0$) and vice versa.

In the absence of interactions, the dashed curves in Fig.~\pref{FIG:expt}{(d)} show that the amount of population transferred is roughly the same for the positive and negative sweeps, with the slight difference stemming from initial condensate momentum of $-0.018\hbar k$.~\cite{Inter-SuppMat}.
The presence of site-dependent interactions leads to a highly asymmetric, direction-dependent response in the population dynamics.
Comparing the positive sweep data (left, blue) to the single particle theory, we find that population begins to transfer earlier and more population is transferred at the ramp's end.
In contrast, the negative sweep data (right, red) exhibits a delay and slightly lower transfer.

The data in Fig.~\pref{FIG:expt}{(d)} are in much better agreement with numerical simulations that incorporate the momentum-space interactions, assuming the inhomogeneous density distribution of the condensate in a harmonic trap~\cite{Stenger-Bragg}.
Specifically, we use a local density approximation, taking a weighted average of simulation curves with different homogeneous mean-field energy $U$ ranging from $0$ to a peak mean-field energy $U_0$~\cite{Inter-SuppMat}.
%For a uniform gas, this shift would relate to a homogeneous mean-field energy $U/\hbar \approx 2\pi\times 1.81$~kHz.
%In our harmonic trap, $U$ relates instead to the average mean-field shift in the trapped gas, with local mean-field energies ranging from zero to $U_0/\hbar \approx 2\pi\times 3.17$~kHz in the central region with peak density~\cite{Stenger-Bragg}.
To perform these simulations, we consider a simplified description of the exact interacting many-body system.
First, we assume that all momentum states occupy the same spatial mode, ignoring effects of spatial separation. 
%Furthermore, we assume that the total particle density is spatially uniform and fixed in time, relating to a uniform mean-field energy $U/\hbar \approx 2\pi \times 1.81$~kHz.
Furthermore, we assume that the allowed momenta values relate to fully distinguishable quantum states.
Thus the momentum-space interaction, which in general is highly non-trivial (depending on the total density and the exact distribution of all site populations), becomes completely site-local such that the interaction between atoms in different momentum states is exactly twice that of atoms occupying the same state.
This approximation is strictly valid in the limit $2U \ll 4 E_R$, and is approximately true for our experimental conditions. Lastly, given that the number of atoms in experiment vastly exceeds the number of sites, we ignore quantum fluctuations and simply represent the condensate wavefunction by appropriately normalized complex amplitudes $\phi_n$ for the various discrete plane-wave momentum states~\cite{Trippenbach-FWM-theory}.

Under these conditions, for the case of a generic single-particle tight-binding model $H^\text{sp}$ (Eq.~\ref{EQ:e00}), the influence of momentum-space interactions may be captured by the nonlinear Schr\"{o}dinger equation
\begin{equation}
i \hbar \dot{\phi}_n = \sum_m H^\text{sp}_{mn} \phi_m + U [2 - |\phi_n|^2 ] \phi_n \ ,
\label{EQ:e0e}
\end{equation}
where $H^\text{sp}_{mn}$ is the matrix element of $H^\text{sp}$ associated with the states $p_m$ and $p_n$. The form of Eq.~\ref{EQ:e0e}, which assumes the normalization condition $\sum_n |\phi_n|^2 = 1$, hints at the effectively attractive, mode-local momentum-space interaction. We note that all interaction terms preserve the individual site populations, as mode mixing is disallowed for the case of elastic, one-dimensional collisions and a uniform density~\cite{Deng-FWM-1999,Trippenbach-FWM-theory,Rolston-NL-2002}, except when considering multiple internal states~\cite{Pertot-10-PRL} or a lattice-modified dispersion~\cite{Gretchen-FWM}. The simulated dynamics for the double-well case, where $H^\text{sp} = \Delta (\tau) \ \hat{c}_{1}^{\dag}\hat{c}_{1} - t(\hat{c}_{0}^{\dag}\hat{c}_{1} + \hat{c}_{1}^{\dag}\hat{c}_{0})$ for time $\tau$, are shown along with the data in Fig.~\pref{FIG:expt}{(d)}. These simulations, which reproduce the observed direction-dependent response, agree much better with the data. 
The lack of oscillatory behavior in the data can be attributed to spatial decoherence between momentum orders. We performed a combined fit of the data from Figs.~\pref{FIG:expt}{(a-d)} to obtain values for tunneling energy $t/\hbar \approx 2\pi\times 390$~Hz, initial condensate momentum $-0.018 \hbar k$, and a peak mean-field energy $U_0/\hbar \approx 2\pi \times 3.17$~kHz of our inhomogeneous density distribution, with an average mean-field energy of $U/\hbar \approx 2\pi\times 1.81$~kHz~\cite{Inter-SuppMat}.

These direction-dependent dynamics can be better understood by considering the adiabatic energy levels of this coupled two-level system~\cite{Inter-SuppMat}, and their projections onto the measured well populations. Without interactions, the Bragg-driven ``tunneling'' leads to an avoided crossing of the adiabatic energy levels (Fig.~\pref{FIG:expt}{(e)}), allowing for efficient transfer in the case of a very slow ramp of the intersite bias (Fig.~\pref{FIG:expt}{(f)}). With interactions (Fig.~\pref{FIG:expt}{(g,h)}), a swallow-tail-like looped structure appears in the energy levels~\cite{Fallani-instable,GORDON,Koller-Looped}, preventing the adiabatic transfer of population. In particular, for a positive sweep one overshoots the transfer from $L$ to $R$, and forced tunneling between the two energy branches leads to inhibited transfer and self-trapping for any finite ramp rate~\cite{GORDON}. This swallow-tail structure is similar to that found in coupled angular momentum states of annular Bose gases~\cite{Eckel-Hysteresis-2014,Ryu-SQUID}, where similar hysteretic behavior has been observed.

\begin{figure}[t!]
\includegraphics{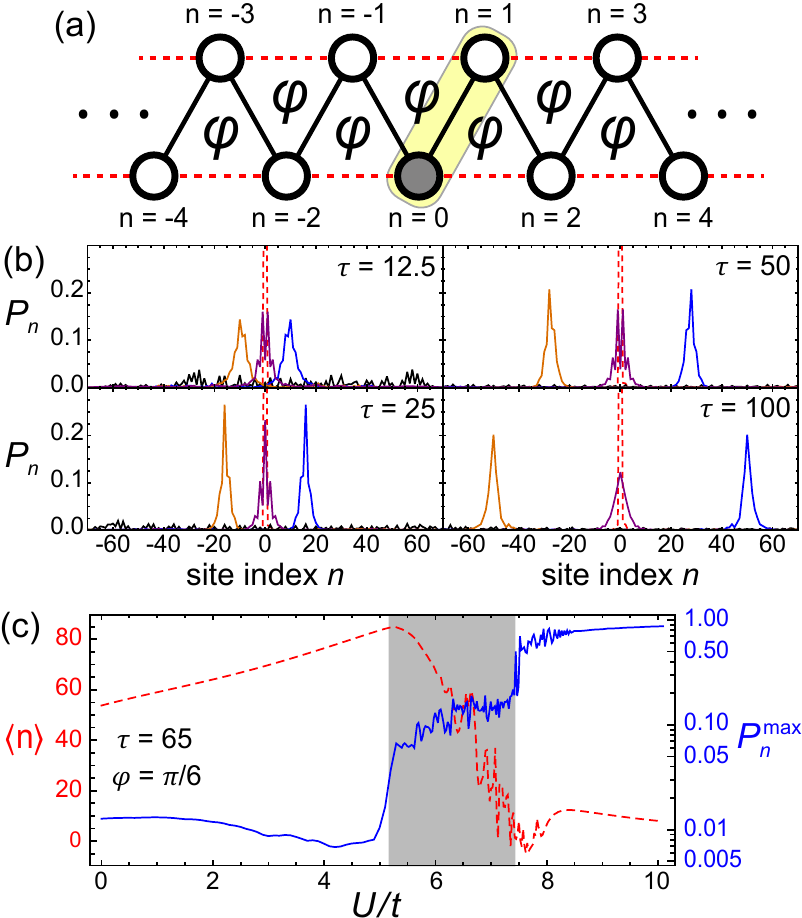}
\caption{\label{FIG:zigzag}
\textbf{Interaction effects in a topological lattice.}
(\textbf{a})~Cartoon depiction of atoms on a zigzag lattice with uniform magnetic flux $\varphi$ and population initialized at the central site $n=0$ (gray). Tunneling links realized by first- (solid black) and second-order (dashed red) Bragg transitions have uniform amplitude $t$. Yellow shaded region indicates two-site unit cell.
(\textbf{b})~Site population distributions for evolution times $\tau = \{12.5, 25, 50, 100\} \, \hbar/t$, shown for several combinations of interaction-to-tunneling ratios ($U/t$) and flux values ($\varphi$) on a 401-site lattice. Solid black denotes $(U/t , \varphi ) = (0 , \pi/6)$, rightmost solid blue for $(7.2 , \pi/6)$, dashed red for $(12 , \pi/6)$, solid purple for $(7.2 , 0)$, and leftmost solid orange for $(7.2 , -\pi/6)$.
(\textbf{c})~Average site position $\langle n \rangle$ (red dashed line; left vertical axis with linear scale) and population in the most-populated site $P_n^{\mathrm{max}}$ (blue solid line; right vertical axis with logarithmic scale) versus $U/t$. The simulations are shown for $\varphi = \pi/6$ after an evolution time $\tau = 65 \, \hbar / t$ on an 801-site lattice, with population never reaching the boundaries. The shaded gray region relates to chiral soliton stability.
}
\end{figure}

These interactions may also allow the study of symmetry breaking~\cite{Trenkwalder-ParSymmBreak-2016} and two-mode entanglement~\cite{RaghavanBigelow} in momentum-space double wells. In particular, the generation of squeezed momentum states through the effectively mode-local interactions could lead to practical advances in inertial sensing~\cite{Inter-SuppMat}. Looking beyond the rich double well system~\cite{Albiez-Oberth-2005,Levy-ACDC-2007,Leblanc-Joseph,Eckel-Hysteresis-2014,Trenkwalder-ParSymmBreak-2016,Chang-spinor-josephson,Tomko-Oberth-2017}, MSLs offer unique capabilities for engineering multiply-connected lattice geometries.
Here we explore the influence of interactions on the particle dynamics in a topological zigzag lattice (Fig.~\pref{FIG:zigzag}), where artificial magnetic fluxes play a nontrivial role~\cite{An-Zigzag}.

We consider a zigzag lattice comprised of uniform nearest and next-nearest neighbor tunneling $t_{n,1} = t_{n,2} \equiv t$ and a uniform magnetic flux $\varphi$ (Fig.~\pref{FIG:zigzag}{(a)}). We examine the case of population initialized to a single, central mode ($n=0$), exploring dynamics following a tunneling quench. The distributions of normalized site populations $P_n$ at various evolution times $\tau$ are shown in Fig.~\pref{FIG:zigzag}{(b)}. Considering a positive flux value of $\pi/6$, dramatically different behavior is found for the cases of no ($U/t = 0$), moderate ($U/t = 7.2$), and strong ($U/t = 12$) interactions. With no interactions, chiral currents are present, but with a rapid ballistic spreading of the atomic distribution. Moderate interactions stabilize the atomic distribution, leading to soliton- or breather-like states~\cite{Tromb-Breather}. For strong interactions, the atoms remain effectively localized at $n = 0$ due to nonlinear self-trapping.

To gain more insight into the general behavior for non-zero fluxes and varied interactions, we plot in Fig.~\pref{FIG:zigzag}{(c)} the dependence of the average site position $\langle n \rangle$ and the population in the most-populated site $P_n^\text{max}$ on the interaction-to-tunneling energy ratio $U/t$, for the case of a $\pi/6$ flux following a duration $\tau = 65 \, \hbar/t$.
For weaker interaction strengths $U/t \lesssim 5$, the initially localized wavepacket becomes highly delocalized (low $P_n^\text{max}$), while on average the population moves in a chiral fashion, reflecting an underlying spin-momentum locking of the topological lattice model~\cite{An-Zigzag}.
In the intermediate regime ($5 \lesssim U/t \lesssim 7.5$), dynamics relating to self-stabilized chiral solitons can be found.
Over the fixed evolution time ($65 \, \hbar/t$), these solitons travel less far as they become more and more ``massive'' with increasing $U/t$. This eventually gives way to full self-trapping and the inhibition of chiral currents for $U/t \gtrsim 7.5$, with population remaining largely localized to the initially-populated site.

The chiral behavior observed for weak interactions stems from the presence of spin-momentum locking in the single-particle band structure, where an effective ``spin'' degree of freedom relates to the two sites of the zigzag lattice unit cell~\cite{Ani-synthz-zigzag,An-Zigzag} (shaded in Fig.~\pref{FIG:zigzag}{(a)}).
The emergence of non-dispersing chiral solitons can be understood in terms of interaction-driven hybridization~\cite{Tai-FluxLadderInt} of the two energy bands in the system. Stability is found as the interaction energy $U$ starts to exceed the width of the lower energy band ($4t$), and a complete self-trapping ensues when the interactions dominate over the combined band width ($6t$).
The collective chiral behavior of the atoms under intermediate interactions is of fundamental interest to understanding how emergent behavior can arise from the interplay of interactions and lattice topology in kinetically frustrated systems~\cite{DiamondLattice-2016,Vito-Flatband-14,Ehud-FlatBand-10,FlatBand-Nonlinear,Ani-synthz-zigzag}, and may also find use in inertial sensing applications.

We have shown that atomic interaction effects can lead to significantly modified particle dynamics in engineered MSLs. These effects result from a combination of momentum-independent scattering and the exchange symmetry of pairwise collisions for bosonic atoms. We have observed the onset of nonlinear self-trapping through the population dynamics of atoms in a momentum-space double well with controlled site-to-site energy bias. In considering the role of momentum-space interactions in the types of topological lattices that may be engineered by MSL techniques~\cite{An-Zigzag}, we show that stable chiral solitons can emerge. Highly tunable MSLs should also enable the exploration into the interplay of long-ranged momentum-space interactions and disorder-driven localization effects~\cite{aleiner:finite_temperature_disorder_2010,Deissler-DisorderWithInteractions-2010}.

%\bibliographystyle{apsrev4-1}
%\bibliography{BibsInt}

\begin{thebibliography}{59}%
\makeatletter
\providecommand \@ifxundefined [1]{%
 \@ifx{#1\undefined}
}%
\providecommand \@ifnum [1]{%
 \ifnum #1\expandafter \@firstoftwo
 \else \expandafter \@secondoftwo
 \fi
}%
\providecommand \@ifx [1]{%
 \ifx #1\expandafter \@firstoftwo
 \else \expandafter \@secondoftwo
 \fi
}%
\providecommand \natexlab [1]{#1}%
\providecommand \enquote  [1]{``#1''}%
\providecommand \bibnamefont  [1]{#1}%
\providecommand \bibfnamefont [1]{#1}%
\providecommand \citenamefont [1]{#1}%
\providecommand \href@noop [0]{\@secondoftwo}%
\providecommand \href [0]{\begingroup \@sanitize@url \@href}%
\providecommand \@href[1]{\@@startlink{#1}\@@href}%
\providecommand \@@href[1]{\endgroup#1\@@endlink}%
\providecommand \@sanitize@url [0]{\catcode `\\12\catcode `\$12\catcode
  `\&12\catcode `\#12\catcode `\^12\catcode `\_12\catcode `\%12\relax}%
\providecommand \@@startlink[1]{}%
\providecommand \@@endlink[0]{}%
\providecommand \url  [0]{\begingroup\@sanitize@url \@url }%
\providecommand \@url [1]{\endgroup\@href {#1}{\urlprefix }}%
\providecommand \urlprefix  [0]{URL }%
\providecommand \Eprint [0]{\href }%
\providecommand \doibase [0]{http://dx.doi.org/}%
\providecommand \selectlanguage [0]{\@gobble}%
\providecommand \bibinfo  [0]{\@secondoftwo}%
\providecommand \bibfield  [0]{\@secondoftwo}%
\providecommand \translation [1]{[#1]}%
\providecommand \BibitemOpen [0]{}%
\providecommand \bibitemStop [0]{}%
\providecommand \bibitemNoStop [0]{.\EOS\space}%
\providecommand \EOS [0]{\spacefactor3000\relax}%
\providecommand \BibitemShut  [1]{\csname bibitem#1\endcsname}%
\let\auto@bib@innerbib\@empty
%</preamble>
\bibitem [{\citenamefont {Bloch}\ \emph {et~al.}(2008)\citenamefont {Bloch},
  \citenamefont {Dalibard},\ and\ \citenamefont {Zwerger}}]{Bloch-RMP08}%
  \BibitemOpen
  \bibfield  {author} {\bibinfo {author} {\bibfnamefont {I.}~\bibnamefont
  {Bloch}}, \bibinfo {author} {\bibfnamefont {J.}~\bibnamefont {Dalibard}}, \
  and\ \bibinfo {author} {\bibfnamefont {W.}~\bibnamefont {Zwerger}},\ }\href
  {\doibase 10.1103/RevModPhys.80.885} {\bibfield  {journal} {\bibinfo
  {journal} {Rev. Mod. Phys.}\ }\textbf {\bibinfo {volume} {80}},\ \bibinfo
  {pages} {885} (\bibinfo {year} {2008})}\BibitemShut {NoStop}%
\bibitem [{\citenamefont {Bloch}\ \emph {et~al.}(2012)\citenamefont {Bloch},
  \citenamefont {Dalibard},\ and\ \citenamefont
  {Nascimb\`{e}ne}}]{AtomsRev-NatPhys-2012}%
  \BibitemOpen
  \bibfield  {author} {\bibinfo {author} {\bibfnamefont {I.}~\bibnamefont
  {Bloch}}, \bibinfo {author} {\bibfnamefont {J.}~\bibnamefont {Dalibard}}, \
  and\ \bibinfo {author} {\bibfnamefont {S.}~\bibnamefont {Nascimb\`{e}ne}},\
  }\href {\doibase 10.1038/nphys2259} {\bibfield  {journal} {\bibinfo
  {journal} {Nat. Phys.}\ }\textbf {\bibinfo {volume} {8}},\ \bibinfo {pages}
  {267} (\bibinfo {year} {2012})}\BibitemShut {NoStop}%
\bibitem [{\citenamefont {Boada}\ \emph {et~al.}(2012)\citenamefont {Boada},
  \citenamefont {Celi}, \citenamefont {Latorre},\ and\ \citenamefont
  {Lewenstein}}]{Boada-Synth}%
  \BibitemOpen
  \bibfield  {author} {\bibinfo {author} {\bibfnamefont {O.}~\bibnamefont
  {Boada}}, \bibinfo {author} {\bibfnamefont {A.}~\bibnamefont {Celi}},
  \bibinfo {author} {\bibfnamefont {J.~I.}\ \bibnamefont {Latorre}}, \ and\
  \bibinfo {author} {\bibfnamefont {M.}~\bibnamefont {Lewenstein}},\ }\href
  {\doibase 10.1103/PhysRevLett.108.133001} {\bibfield  {journal} {\bibinfo
  {journal} {Phys. Rev. Lett.}\ }\textbf {\bibinfo {volume} {108}},\ \bibinfo
  {pages} {133001} (\bibinfo {year} {2012})}\BibitemShut {NoStop}%
\bibitem [{\citenamefont {Celi}\ \emph {et~al.}(2014)\citenamefont {Celi},
  \citenamefont {Massignan}, \citenamefont {Ruseckas}, \citenamefont {Goldman},
  \citenamefont {Spielman}, \citenamefont {Juzeli\ifmmode~\bar{u}\else
  \={u}\fi{}nas},\ and\ \citenamefont {Lewenstein}}]{Celi-ArtificialDim}%
  \BibitemOpen
  \bibfield  {author} {\bibinfo {author} {\bibfnamefont {A.}~\bibnamefont
  {Celi}}, \bibinfo {author} {\bibfnamefont {P.}~\bibnamefont {Massignan}},
  \bibinfo {author} {\bibfnamefont {J.}~\bibnamefont {Ruseckas}}, \bibinfo
  {author} {\bibfnamefont {N.}~\bibnamefont {Goldman}}, \bibinfo {author}
  {\bibfnamefont {I.~B.}\ \bibnamefont {Spielman}}, \bibinfo {author}
  {\bibfnamefont {G.}~\bibnamefont {Juzeli\ifmmode~\bar{u}\else
  \={u}\fi{}nas}}, \ and\ \bibinfo {author} {\bibfnamefont {M.}~\bibnamefont
  {Lewenstein}},\ }\href {\doibase 10.1103/PhysRevLett.112.043001} {\bibfield
  {journal} {\bibinfo  {journal} {Phys. Rev. Lett.}\ }\textbf {\bibinfo
  {volume} {112}},\ \bibinfo {pages} {043001} (\bibinfo {year}
  {2014})}\BibitemShut {NoStop}%
\bibitem [{\citenamefont {Price}\ \emph {et~al.}(2017)\citenamefont {Price},
  \citenamefont {Ozawa},\ and\ \citenamefont {Goldman}}]{NateGold-TrapShake}%
  \BibitemOpen
  \bibfield  {author} {\bibinfo {author} {\bibfnamefont {H.~M.}\ \bibnamefont
  {Price}}, \bibinfo {author} {\bibfnamefont {T.}~\bibnamefont {Ozawa}}, \ and\
  \bibinfo {author} {\bibfnamefont {N.}~\bibnamefont {Goldman}},\ }\href
  {\doibase 10.1103/PhysRevA.95.023607} {\bibfield  {journal} {\bibinfo
  {journal} {Phys. Rev. A}\ }\textbf {\bibinfo {volume} {95}},\ \bibinfo
  {pages} {023607} (\bibinfo {year} {2017})}\BibitemShut {NoStop}%
\bibitem [{\citenamefont {Wall}\ \emph {et~al.}(2016)\citenamefont {Wall},
  \citenamefont {Koller}, \citenamefont {Li}, \citenamefont {Zhang},
  \citenamefont {Cooper}, \citenamefont {Ye},\ and\ \citenamefont
  {Rey}}]{Wall-SpinOrb}%
  \BibitemOpen
  \bibfield  {author} {\bibinfo {author} {\bibfnamefont {M.~L.}\ \bibnamefont
  {Wall}}, \bibinfo {author} {\bibfnamefont {A.~P.}\ \bibnamefont {Koller}},
  \bibinfo {author} {\bibfnamefont {S.}~\bibnamefont {Li}}, \bibinfo {author}
  {\bibfnamefont {X.}~\bibnamefont {Zhang}}, \bibinfo {author} {\bibfnamefont
  {N.~R.}\ \bibnamefont {Cooper}}, \bibinfo {author} {\bibfnamefont
  {J.}~\bibnamefont {Ye}}, \ and\ \bibinfo {author} {\bibfnamefont {A.~M.}\
  \bibnamefont {Rey}},\ }\href {\doibase 10.1103/PhysRevLett.116.035301}
  {\bibfield  {journal} {\bibinfo  {journal} {Phys. Rev. Lett.}\ }\textbf
  {\bibinfo {volume} {116}},\ \bibinfo {pages} {035301} (\bibinfo {year}
  {2016})}\BibitemShut {NoStop}%
\bibitem [{\citenamefont {Gadway}(2015)}]{Gadway-KSPACE}%
  \BibitemOpen
  \bibfield  {author} {\bibinfo {author} {\bibfnamefont {B.}~\bibnamefont
  {Gadway}},\ }\href {\doibase 10.1103/PhysRevA.92.043606} {\bibfield
  {journal} {\bibinfo  {journal} {Phys. Rev. A}\ }\textbf {\bibinfo {volume}
  {92}},\ \bibinfo {pages} {043606} (\bibinfo {year} {2015})}\BibitemShut
  {NoStop}%
\bibitem [{\citenamefont {Meier}\ \emph
  {et~al.}(2016{\natexlab{a}})\citenamefont {Meier}, \citenamefont {An},\ and\
  \citenamefont {Gadway}}]{Meier-AtomOptics}%
  \BibitemOpen
  \bibfield  {author} {\bibinfo {author} {\bibfnamefont {E.~J.}\ \bibnamefont
  {Meier}}, \bibinfo {author} {\bibfnamefont {F.~A.}\ \bibnamefont {An}}, \
  and\ \bibinfo {author} {\bibfnamefont {B.}~\bibnamefont {Gadway}},\ }\href
  {\doibase 10.1103/PhysRevA.93.051602} {\bibfield  {journal} {\bibinfo
  {journal} {Phys. Rev. A}\ }\textbf {\bibinfo {volume} {93}},\ \bibinfo
  {pages} {051602} (\bibinfo {year} {2016}{\natexlab{a}})}\BibitemShut
  {NoStop}%
\bibitem [{\citenamefont {Meier}\ \emph
  {et~al.}(2016{\natexlab{b}})\citenamefont {Meier}, \citenamefont {An},\ and\
  \citenamefont {Gadway}}]{Meier-SSH}%
  \BibitemOpen
  \bibfield  {author} {\bibinfo {author} {\bibfnamefont {E.~J.}\ \bibnamefont
  {Meier}}, \bibinfo {author} {\bibfnamefont {F.~A.}\ \bibnamefont {An}}, \
  and\ \bibinfo {author} {\bibfnamefont {B.}~\bibnamefont {Gadway}},\ }\href
  {\doibase 10.1038/ncomms13986} {\bibfield  {journal} {\bibinfo  {journal}
  {Nat. Commun.}\ }\textbf {\bibinfo {volume} {7}},\ \bibinfo {pages} {13986}
  (\bibinfo {year} {2016}{\natexlab{b}})}\BibitemShut {NoStop}%
\bibitem [{\citenamefont {An}\ \emph {et~al.}(2017{\natexlab{a}})\citenamefont
  {An}, \citenamefont {Meier},\ and\ \citenamefont {Gadway}}]{Alex-2Dchiral}%
  \BibitemOpen
  \bibfield  {author} {\bibinfo {author} {\bibfnamefont {F.~A.}\ \bibnamefont
  {An}}, \bibinfo {author} {\bibfnamefont {E.~J.}\ \bibnamefont {Meier}}, \
  and\ \bibinfo {author} {\bibfnamefont {B.}~\bibnamefont {Gadway}},\ }\href
  {\doibase 10.1126/sciadv.1602685} {\bibfield  {journal} {\bibinfo  {journal}
  {Sci. Adv.}\ }\textbf {\bibinfo {volume} {3}},\ \bibinfo {pages} {e1602685}
  (\bibinfo {year} {2017}{\natexlab{a}})}\BibitemShut {NoStop}%
\bibitem [{\citenamefont {An}\ \emph {et~al.}(2017{\natexlab{b}})\citenamefont
  {An}, \citenamefont {Meier},\ and\ \citenamefont {Gadway}}]{Alex-Annealed}%
  \BibitemOpen
  \bibfield  {author} {\bibinfo {author} {\bibfnamefont {F.~A.}\ \bibnamefont
  {An}}, \bibinfo {author} {\bibfnamefont {E.~J.}\ \bibnamefont {Meier}}, \
  and\ \bibinfo {author} {\bibfnamefont {B.}~\bibnamefont {Gadway}},\
  }\href@noop {} {\  (\bibinfo {year} {2017}{\natexlab{b}})},\ \Eprint
  {http://arxiv.org/abs/1701.07493} {arXiv:1701.07493} \BibitemShut {NoStop}%
\bibitem [{\citenamefont {Mancini}\ \emph {et~al.}(2015)\citenamefont
  {Mancini}, \citenamefont {Pagano}, \citenamefont {Cappellini}, \citenamefont
  {Livi}, \citenamefont {Rider}, \citenamefont {Catani}, \citenamefont {Sias},
  \citenamefont {Zoller}, \citenamefont {Inguscio}, \citenamefont {Dalmonte},\
  and\ \citenamefont {Fallani}}]{Fallani-chiral-2015}%
  \BibitemOpen
  \bibfield  {author} {\bibinfo {author} {\bibfnamefont {M.}~\bibnamefont
  {Mancini}}, \bibinfo {author} {\bibfnamefont {G.}~\bibnamefont {Pagano}},
  \bibinfo {author} {\bibfnamefont {G.}~\bibnamefont {Cappellini}}, \bibinfo
  {author} {\bibfnamefont {L.}~\bibnamefont {Livi}}, \bibinfo {author}
  {\bibfnamefont {M.}~\bibnamefont {Rider}}, \bibinfo {author} {\bibfnamefont
  {J.}~\bibnamefont {Catani}}, \bibinfo {author} {\bibfnamefont
  {C.}~\bibnamefont {Sias}}, \bibinfo {author} {\bibfnamefont {P.}~\bibnamefont
  {Zoller}}, \bibinfo {author} {\bibfnamefont {M.}~\bibnamefont {Inguscio}},
  \bibinfo {author} {\bibfnamefont {M.}~\bibnamefont {Dalmonte}}, \ and\
  \bibinfo {author} {\bibfnamefont {L.}~\bibnamefont {Fallani}},\ }\href
  {\doibase 10.1126/science.aaa8736} {\bibfield  {journal} {\bibinfo  {journal}
  {Science}\ }\textbf {\bibinfo {volume} {349}},\ \bibinfo {pages} {1510}
  (\bibinfo {year} {2015})}\BibitemShut {NoStop}%
\bibitem [{\citenamefont {Stuhl}\ \emph {et~al.}(2015)\citenamefont {Stuhl},
  \citenamefont {Lu}, \citenamefont {Aycock}, \citenamefont {Genkina},\ and\
  \citenamefont {Spielman}}]{Stuhl-Edge-2015}%
  \BibitemOpen
  \bibfield  {author} {\bibinfo {author} {\bibfnamefont {B.~K.}\ \bibnamefont
  {Stuhl}}, \bibinfo {author} {\bibfnamefont {H.-I.}\ \bibnamefont {Lu}},
  \bibinfo {author} {\bibfnamefont {L.~M.}\ \bibnamefont {Aycock}}, \bibinfo
  {author} {\bibfnamefont {D.}~\bibnamefont {Genkina}}, \ and\ \bibinfo
  {author} {\bibfnamefont {I.~B.}\ \bibnamefont {Spielman}},\ }\href {\doibase
  10.1126/science.aaa8515} {\bibfield  {journal} {\bibinfo  {journal}
  {Science}\ }\textbf {\bibinfo {volume} {349}},\ \bibinfo {pages} {1514}
  (\bibinfo {year} {2015})}\BibitemShut {NoStop}%
\bibitem [{\citenamefont {Yan}\ \emph {et~al.}(2013)\citenamefont {Yan},
  \citenamefont {Moses}, \citenamefont {Gadway}, \citenamefont {Covey},
  \citenamefont {Hazzard}, \citenamefont {Rey}, \citenamefont {Jin},\ and\
  \citenamefont {Ye}}]{yan2013:dipole-dipole}%
  \BibitemOpen
  \bibfield  {author} {\bibinfo {author} {\bibfnamefont {B.}~\bibnamefont
  {Yan}}, \bibinfo {author} {\bibfnamefont {S.~A.}\ \bibnamefont {Moses}},
  \bibinfo {author} {\bibfnamefont {B.}~\bibnamefont {Gadway}}, \bibinfo
  {author} {\bibfnamefont {J.~P.}\ \bibnamefont {Covey}}, \bibinfo {author}
  {\bibfnamefont {K.~R.~A.}\ \bibnamefont {Hazzard}}, \bibinfo {author}
  {\bibfnamefont {A.~M.}\ \bibnamefont {Rey}}, \bibinfo {author} {\bibfnamefont
  {D.~S.}\ \bibnamefont {Jin}}, \ and\ \bibinfo {author} {\bibfnamefont
  {J.}~\bibnamefont {Ye}},\ }\href {http://dx.doi.org/10.1038/nature12483}
  {\bibfield  {journal} {\bibinfo  {journal} {Nature}\ }\textbf {\bibinfo
  {volume} {501}},\ \bibinfo {pages} {521} (\bibinfo {year}
  {2013})}\BibitemShut {NoStop}%
\bibitem [{\citenamefont {Szameit}\ and\ \citenamefont
  {Nolte}(2010)}]{SzameitReview-2010}%
  \BibitemOpen
  \bibfield  {author} {\bibinfo {author} {\bibfnamefont {A.}~\bibnamefont
  {Szameit}}\ and\ \bibinfo {author} {\bibfnamefont {S.}~\bibnamefont
  {Nolte}},\ }\href {\doibase 10.1088/0953-4075/43/16/163001} {\bibfield
  {journal} {\bibinfo  {journal} {J. Phys. B}\ }\textbf {\bibinfo {volume}
  {43}},\ \bibinfo {pages} {163001} (\bibinfo {year} {2010})}\BibitemShut
  {NoStop}%
\bibitem [{\citenamefont {Aspuru-Guzik}\ and\ \citenamefont
  {Walther}(2012)}]{PhotRev-NatPhys-2012}%
  \BibitemOpen
  \bibfield  {author} {\bibinfo {author} {\bibfnamefont {A.}~\bibnamefont
  {Aspuru-Guzik}}\ and\ \bibinfo {author} {\bibfnamefont {P.}~\bibnamefont
  {Walther}},\ }\href {\doibase doi:10.1038/nphys2253} {\bibfield  {journal}
  {\bibinfo  {journal} {Nat. Phys.}\ }\textbf {\bibinfo {volume} {8}},\
  \bibinfo {pages} {285} (\bibinfo {year} {2012})}\BibitemShut {NoStop}%
\bibitem [{\citenamefont {Sugawa}\ \emph {et~al.}(2016)\citenamefont {Sugawa},
  \citenamefont {Salces-Carcoba}, \citenamefont {Perry}, \citenamefont {Yue},\
  and\ \citenamefont {Spielman}}]{Sugawa-NonAb}%
  \BibitemOpen
  \bibfield  {author} {\bibinfo {author} {\bibfnamefont {S.}~\bibnamefont
  {Sugawa}}, \bibinfo {author} {\bibfnamefont {F.}~\bibnamefont
  {Salces-Carcoba}}, \bibinfo {author} {\bibfnamefont {A.~R.}\ \bibnamefont
  {Perry}}, \bibinfo {author} {\bibfnamefont {Y.}~\bibnamefont {Yue}}, \ and\
  \bibinfo {author} {\bibfnamefont {I.~B.}\ \bibnamefont {Spielman}},\
  }\href@noop {} {\  (\bibinfo {year} {2016})},\ \Eprint
  {http://arxiv.org/abs/1610.06228} {arXiv:1610.06228} \BibitemShut {NoStop}%
\bibitem [{\citenamefont {S\"oding}\ \emph {et~al.}(1998)\citenamefont
  {S\"oding}, \citenamefont {Gu\'ery-Odelin}, \citenamefont {Desbiolles},
  \citenamefont {Ferrari},\ and\ \citenamefont {Dalibard}}]{Soding-relaxation}%
  \BibitemOpen
  \bibfield  {author} {\bibinfo {author} {\bibfnamefont {J.}~\bibnamefont
  {S\"oding}}, \bibinfo {author} {\bibfnamefont {D.}~\bibnamefont
  {Gu\'ery-Odelin}}, \bibinfo {author} {\bibfnamefont {P.}~\bibnamefont
  {Desbiolles}}, \bibinfo {author} {\bibfnamefont {G.}~\bibnamefont {Ferrari}},
  \ and\ \bibinfo {author} {\bibfnamefont {J.}~\bibnamefont {Dalibard}},\
  }\href {\doibase 10.1103/PhysRevLett.80.1869} {\bibfield  {journal} {\bibinfo
   {journal} {Phys. Rev. Lett.}\ }\textbf {\bibinfo {volume} {80}},\ \bibinfo
  {pages} {1869} (\bibinfo {year} {1998})}\BibitemShut {NoStop}%
\bibitem [{\citenamefont {Weiner}\ \emph {et~al.}(1999)\citenamefont {Weiner},
  \citenamefont {Bagnato}, \citenamefont {Zilio},\ and\ \citenamefont
  {Julienne}}]{Weiner-Collisions}%
  \BibitemOpen
  \bibfield  {author} {\bibinfo {author} {\bibfnamefont {J.}~\bibnamefont
  {Weiner}}, \bibinfo {author} {\bibfnamefont {V.~S.}\ \bibnamefont {Bagnato}},
  \bibinfo {author} {\bibfnamefont {S.}~\bibnamefont {Zilio}}, \ and\ \bibinfo
  {author} {\bibfnamefont {P.~S.}\ \bibnamefont {Julienne}},\ }\href {\doibase
  10.1103/RevModPhys.71.1} {\bibfield  {journal} {\bibinfo  {journal} {Rev.
  Mod. Phys.}\ }\textbf {\bibinfo {volume} {71}},\ \bibinfo {pages} {1}
  (\bibinfo {year} {1999})}\BibitemShut {NoStop}%
\bibitem [{\citenamefont {Pagano}\ \emph {et~al.}(2014)\citenamefont {Pagano},
  \citenamefont {Mancini}, \citenamefont {Cappellini}, \citenamefont
  {Lombardi}, \citenamefont {Schafer}, \citenamefont {Hu}, \citenamefont {Liu},
  \citenamefont {Catani}, \citenamefont {Sias}, \citenamefont {Inguscio},\ and\
  \citenamefont {Fallani}}]{Pagano-AlkalineEarth}%
  \BibitemOpen
  \bibfield  {author} {\bibinfo {author} {\bibfnamefont {G.}~\bibnamefont
  {Pagano}}, \bibinfo {author} {\bibfnamefont {M.}~\bibnamefont {Mancini}},
  \bibinfo {author} {\bibfnamefont {G.}~\bibnamefont {Cappellini}}, \bibinfo
  {author} {\bibfnamefont {P.}~\bibnamefont {Lombardi}}, \bibinfo {author}
  {\bibfnamefont {F.}~\bibnamefont {Schafer}}, \bibinfo {author} {\bibfnamefont
  {H.}~\bibnamefont {Hu}}, \bibinfo {author} {\bibfnamefont {X.-J.}\
  \bibnamefont {Liu}}, \bibinfo {author} {\bibfnamefont {J.}~\bibnamefont
  {Catani}}, \bibinfo {author} {\bibfnamefont {C.}~\bibnamefont {Sias}},
  \bibinfo {author} {\bibfnamefont {M.}~\bibnamefont {Inguscio}}, \ and\
  \bibinfo {author} {\bibfnamefont {L.}~\bibnamefont {Fallani}},\ }\href
  {http://dx.doi.org/10.1038/nphys2878} {\bibfield  {journal} {\bibinfo
  {journal} {Nat. Phys.}\ }\textbf {\bibinfo {volume} {10}},\ \bibinfo {pages}
  {198} (\bibinfo {year} {2014})}\BibitemShut {NoStop}%
\bibitem [{\citenamefont {Kozuma}\ \emph {et~al.}(1999)\citenamefont {Kozuma},
  \citenamefont {Deng}, \citenamefont {Hagley}, \citenamefont {Wen},
  \citenamefont {Lutwak}, \citenamefont {Helmerson}, \citenamefont {Rolston},\
  and\ \citenamefont {Phillips}}]{Kozuma-Bragg}%
  \BibitemOpen
  \bibfield  {author} {\bibinfo {author} {\bibfnamefont {M.}~\bibnamefont
  {Kozuma}}, \bibinfo {author} {\bibfnamefont {L.}~\bibnamefont {Deng}},
  \bibinfo {author} {\bibfnamefont {E.~W.}\ \bibnamefont {Hagley}}, \bibinfo
  {author} {\bibfnamefont {J.}~\bibnamefont {Wen}}, \bibinfo {author}
  {\bibfnamefont {R.}~\bibnamefont {Lutwak}}, \bibinfo {author} {\bibfnamefont
  {K.}~\bibnamefont {Helmerson}}, \bibinfo {author} {\bibfnamefont {S.~L.}\
  \bibnamefont {Rolston}}, \ and\ \bibinfo {author} {\bibfnamefont {W.~D.}\
  \bibnamefont {Phillips}},\ }\href {\doibase 10.1103/PhysRevLett.82.871}
  {\bibfield  {journal} {\bibinfo  {journal} {Phys. Rev. Lett.}\ }\textbf
  {\bibinfo {volume} {82}},\ \bibinfo {pages} {871} (\bibinfo {year}
  {1999})}\BibitemShut {NoStop}%
\bibitem [{\citenamefont {Greiner}\ \emph {et~al.}(2002)\citenamefont
  {Greiner}, \citenamefont {Mandel}, \citenamefont {Esslinger}, \citenamefont
  {Hansch},\ and\ \citenamefont {Bloch}}]{Greiner-SF-MI-2002}%
  \BibitemOpen
  \bibfield  {author} {\bibinfo {author} {\bibfnamefont {M.}~\bibnamefont
  {Greiner}}, \bibinfo {author} {\bibfnamefont {O.}~\bibnamefont {Mandel}},
  \bibinfo {author} {\bibfnamefont {T.}~\bibnamefont {Esslinger}}, \bibinfo
  {author} {\bibfnamefont {T.~W.}\ \bibnamefont {Hansch}}, \ and\ \bibinfo
  {author} {\bibfnamefont {I.}~\bibnamefont {Bloch}},\ }\href {\doibase
  10.1038/415039a} {\bibfield  {journal} {\bibinfo  {journal} {Nature}\
  }\textbf {\bibinfo {volume} {415}},\ \bibinfo {pages} {39} (\bibinfo {year}
  {2002})}\BibitemShut {NoStop}%
\bibitem [{\citenamefont {Mazurenko}\ \emph {et~al.}(2017)\citenamefont
  {Mazurenko}, \citenamefont {Chiu}, \citenamefont {Ji}, \citenamefont
  {Parsons}, \citenamefont {Kan\'{a}sz-Nagy}, \citenamefont {Schmidt},
  \citenamefont {Grusdt}, \citenamefont {Demler}, \citenamefont {Greif},\ and\
  \citenamefont {Greiner}}]{Greiner-AFM}%
  \BibitemOpen
  \bibfield  {author} {\bibinfo {author} {\bibfnamefont {A.}~\bibnamefont
  {Mazurenko}}, \bibinfo {author} {\bibfnamefont {C.~S.}\ \bibnamefont {Chiu}},
  \bibinfo {author} {\bibfnamefont {G.}~\bibnamefont {Ji}}, \bibinfo {author}
  {\bibfnamefont {M.~F.}\ \bibnamefont {Parsons}}, \bibinfo {author}
  {\bibfnamefont {M.}~\bibnamefont {Kan\'{a}sz-Nagy}}, \bibinfo {author}
  {\bibfnamefont {R.}~\bibnamefont {Schmidt}}, \bibinfo {author} {\bibfnamefont
  {F.}~\bibnamefont {Grusdt}}, \bibinfo {author} {\bibfnamefont
  {E.}~\bibnamefont {Demler}}, \bibinfo {author} {\bibfnamefont
  {D.}~\bibnamefont {Greif}}, \ and\ \bibinfo {author} {\bibfnamefont
  {M.}~\bibnamefont {Greiner}},\ }\href {\doibase 10.1038/nature22362}
  {\bibfield  {journal} {\bibinfo  {journal} {Nature}\ }\textbf {\bibinfo
  {volume} {545}},\ \bibinfo {pages} {462} (\bibinfo {year}
  {2017})}\BibitemShut {NoStop}%
\bibitem [{\citenamefont {Ozeri}\ \emph {et~al.}(2005)\citenamefont {Ozeri},
  \citenamefont {Katz}, \citenamefont {Steinhauer},\ and\ \citenamefont
  {Davidson}}]{Ozeri-Bog-RMP}%
  \BibitemOpen
  \bibfield  {author} {\bibinfo {author} {\bibfnamefont {R.}~\bibnamefont
  {Ozeri}}, \bibinfo {author} {\bibfnamefont {N.}~\bibnamefont {Katz}},
  \bibinfo {author} {\bibfnamefont {J.}~\bibnamefont {Steinhauer}}, \ and\
  \bibinfo {author} {\bibfnamefont {N.}~\bibnamefont {Davidson}},\ }\href
  {\doibase 10.1103/RevModPhys.77.187} {\bibfield  {journal} {\bibinfo
  {journal} {Rev. Mod. Phys.}\ }\textbf {\bibinfo {volume} {77}},\ \bibinfo
  {pages} {187} (\bibinfo {year} {2005})}\BibitemShut {NoStop}%
\bibitem [{\citenamefont {Anderlini}\ \emph {et~al.}(2007)\citenamefont
  {Anderlini}, \citenamefont {Lee}, \citenamefont {Brown}, \citenamefont
  {Sebby-Strabley}, \citenamefont {Phillips},\ and\ \citenamefont
  {Porto}}]{exch}%
  \BibitemOpen
  \bibfield  {author} {\bibinfo {author} {\bibfnamefont {M.}~\bibnamefont
  {Anderlini}}, \bibinfo {author} {\bibfnamefont {P.~J.}\ \bibnamefont {Lee}},
  \bibinfo {author} {\bibfnamefont {B.~L.}\ \bibnamefont {Brown}}, \bibinfo
  {author} {\bibfnamefont {J.}~\bibnamefont {Sebby-Strabley}}, \bibinfo
  {author} {\bibfnamefont {W.~D.}\ \bibnamefont {Phillips}}, \ and\ \bibinfo
  {author} {\bibfnamefont {J.~V.}\ \bibnamefont {Porto}},\ }\href {\doibase
  10.1038/nature06011} {\bibfield  {journal} {\bibinfo  {journal} {Nature}\
  }\textbf {\bibinfo {volume} {448}},\ \bibinfo {pages} {452} (\bibinfo {year}
  {2007})}\BibitemShut {NoStop}%
\bibitem [{\citenamefont {Kaufman}\ \emph {et~al.}(2015)\citenamefont
  {Kaufman}, \citenamefont {Lester}, \citenamefont {Foss-Feig}, \citenamefont
  {Wall}, \citenamefont {Rey},\ and\ \citenamefont {Regal}}]{Kaufman-Entang}%
  \BibitemOpen
  \bibfield  {author} {\bibinfo {author} {\bibfnamefont {A.~M.}\ \bibnamefont
  {Kaufman}}, \bibinfo {author} {\bibfnamefont {B.~J.}\ \bibnamefont {Lester}},
  \bibinfo {author} {\bibfnamefont {M.}~\bibnamefont {Foss-Feig}}, \bibinfo
  {author} {\bibfnamefont {M.~L.}\ \bibnamefont {Wall}}, \bibinfo {author}
  {\bibfnamefont {A.~M.}\ \bibnamefont {Rey}}, \ and\ \bibinfo {author}
  {\bibfnamefont {C.~A.}\ \bibnamefont {Regal}},\ }\href {\doibase
  10.1038/nature16073} {\bibfield  {journal} {\bibinfo  {journal} {Nature}\
  }\textbf {\bibinfo {volume} {527}},\ \bibinfo {pages} {208} (\bibinfo {year}
  {2015})}\BibitemShut {NoStop}%
\bibitem [{\citenamefont {Larson}\ \emph {et~al.}(2009)\citenamefont {Larson},
  \citenamefont {Collin},\ and\ \citenamefont
  {Martikainen}}]{Larson-MultibandBosons}%
  \BibitemOpen
  \bibfield  {author} {\bibinfo {author} {\bibfnamefont {J.}~\bibnamefont
  {Larson}}, \bibinfo {author} {\bibfnamefont {A.}~\bibnamefont {Collin}}, \
  and\ \bibinfo {author} {\bibfnamefont {J.-P.}\ \bibnamefont {Martikainen}},\
  }\href {\doibase 10.1103/PhysRevA.79.033603} {\bibfield  {journal} {\bibinfo
  {journal} {Phys. Rev. A}\ }\textbf {\bibinfo {volume} {79}},\ \bibinfo
  {pages} {033603} (\bibinfo {year} {2009})}\BibitemShut {NoStop}%
\bibitem [{\citenamefont {Smith}\ \emph {et~al.}(2011)\citenamefont {Smith},
  \citenamefont {Campbell}, \citenamefont {Tammuz},\ and\ \citenamefont
  {Hadzibabic}}]{Smith-Interactions-BEC}%
  \BibitemOpen
  \bibfield  {author} {\bibinfo {author} {\bibfnamefont {R.~P.}\ \bibnamefont
  {Smith}}, \bibinfo {author} {\bibfnamefont {R.~L.~D.}\ \bibnamefont
  {Campbell}}, \bibinfo {author} {\bibfnamefont {N.}~\bibnamefont {Tammuz}}, \
  and\ \bibinfo {author} {\bibfnamefont {Z.}~\bibnamefont {Hadzibabic}},\
  }\href {\doibase 10.1103/PhysRevLett.106.250403} {\bibfield  {journal}
  {\bibinfo  {journal} {Phys. Rev. Lett.}\ }\textbf {\bibinfo {volume} {106}},\
  \bibinfo {pages} {250403} (\bibinfo {year} {2011})}\BibitemShut {NoStop}%
\bibitem [{\citenamefont {Griffin}\ and\ \citenamefont
  {Zaremba}(1997)}]{SecondSound}%
  \BibitemOpen
  \bibfield  {author} {\bibinfo {author} {\bibfnamefont {A.}~\bibnamefont
  {Griffin}}\ and\ \bibinfo {author} {\bibfnamefont {E.}~\bibnamefont
  {Zaremba}},\ }\href {\doibase 10.1103/PhysRevA.56.4839} {\bibfield  {journal}
  {\bibinfo  {journal} {Phys. Rev. A}\ }\textbf {\bibinfo {volume} {56}},\
  \bibinfo {pages} {4839} (\bibinfo {year} {1997})}\BibitemShut {NoStop}%
\bibitem [{\citenamefont {Stenger}\ \emph {et~al.}(1999)\citenamefont
  {Stenger}, \citenamefont {Inouye}, \citenamefont {Chikkatur}, \citenamefont
  {Stamper-Kurn}, \citenamefont {Pritchard},\ and\ \citenamefont
  {Ketterle}}]{Stenger-Bragg}%
  \BibitemOpen
  \bibfield  {author} {\bibinfo {author} {\bibfnamefont {J.}~\bibnamefont
  {Stenger}}, \bibinfo {author} {\bibfnamefont {S.}~\bibnamefont {Inouye}},
  \bibinfo {author} {\bibfnamefont {A.~P.}\ \bibnamefont {Chikkatur}}, \bibinfo
  {author} {\bibfnamefont {D.~M.}\ \bibnamefont {Stamper-Kurn}}, \bibinfo
  {author} {\bibfnamefont {D.~E.}\ \bibnamefont {Pritchard}}, \ and\ \bibinfo
  {author} {\bibfnamefont {W.}~\bibnamefont {Ketterle}},\ }\href {\doibase
  10.1103/PhysRevLett.82.4569} {\bibfield  {journal} {\bibinfo  {journal}
  {Phys. Rev. Lett.}\ }\textbf {\bibinfo {volume} {82}},\ \bibinfo {pages}
  {4569} (\bibinfo {year} {1999})}\BibitemShut {NoStop}%
\bibitem [{\citenamefont {Lopes}\ \emph {et~al.}(2017)\citenamefont {Lopes},
  \citenamefont {Eigen}, \citenamefont {Barker}, \citenamefont {Viebahn},
  \citenamefont {de~Saint-Vincent}, \citenamefont {Navon}, \citenamefont
  {Hadzibabic},\ and\ \citenamefont {Smith}}]{Hadzibabic-Bragg-Strong}%
  \BibitemOpen
  \bibfield  {author} {\bibinfo {author} {\bibfnamefont {R.}~\bibnamefont
  {Lopes}}, \bibinfo {author} {\bibfnamefont {C.}~\bibnamefont {Eigen}},
  \bibinfo {author} {\bibfnamefont {A.}~\bibnamefont {Barker}}, \bibinfo
  {author} {\bibfnamefont {K.~G.~H.}\ \bibnamefont {Viebahn}}, \bibinfo
  {author} {\bibfnamefont {M.~R.}\ \bibnamefont {de~Saint-Vincent}}, \bibinfo
  {author} {\bibfnamefont {N.}~\bibnamefont {Navon}}, \bibinfo {author}
  {\bibfnamefont {Z.}~\bibnamefont {Hadzibabic}}, \ and\ \bibinfo {author}
  {\bibfnamefont {R.~P.}\ \bibnamefont {Smith}},\ }\href@noop {} {\  (\bibinfo
  {year} {2017})},\ \Eprint {http://arxiv.org/abs/1702.02935}
  {arXiv:1702.02935} \BibitemShut {NoStop}%
\bibitem [{Int()}]{Inter-SuppMat}%
  \BibitemOpen
  \href@noop {} {}\bibinfo {note} {See Supplemental Material for details on the
  fitting procedures, adiabatic energy level calculations, and prospects for
  momentum-space squeezing.}\BibitemShut {Stop}%
\bibitem [{\citenamefont {Trippenbach}\ \emph {et~al.}(2000)\citenamefont
  {Trippenbach}, \citenamefont {Band},\ and\ \citenamefont
  {Julienne}}]{Trippenbach-FWM-theory}%
  \BibitemOpen
  \bibfield  {author} {\bibinfo {author} {\bibfnamefont {M.}~\bibnamefont
  {Trippenbach}}, \bibinfo {author} {\bibfnamefont {Y.~B.}\ \bibnamefont
  {Band}}, \ and\ \bibinfo {author} {\bibfnamefont {P.~S.}\ \bibnamefont
  {Julienne}},\ }\href {\doibase 10.1103/PhysRevA.62.023608} {\bibfield
  {journal} {\bibinfo  {journal} {Phys. Rev. A}\ }\textbf {\bibinfo {volume}
  {62}},\ \bibinfo {pages} {023608} (\bibinfo {year} {2000})}\BibitemShut
  {NoStop}%
\bibitem [{\citenamefont {Deng}\ \emph {et~al.}(1999)\citenamefont {Deng},
  \citenamefont {Hagley}, \citenamefont {Wen}, \citenamefont {Trippenbach},
  \citenamefont {Band}, \citenamefont {Julienne}, \citenamefont {Simsarian},
  \citenamefont {Helmerson}, \citenamefont {Rolston},\ and\ \citenamefont
  {Phillips}}]{Deng-FWM-1999}%
  \BibitemOpen
  \bibfield  {author} {\bibinfo {author} {\bibfnamefont {L.}~\bibnamefont
  {Deng}}, \bibinfo {author} {\bibfnamefont {E.~W.}\ \bibnamefont {Hagley}},
  \bibinfo {author} {\bibfnamefont {J.}~\bibnamefont {Wen}}, \bibinfo {author}
  {\bibfnamefont {M.}~\bibnamefont {Trippenbach}}, \bibinfo {author}
  {\bibfnamefont {Y.}~\bibnamefont {Band}}, \bibinfo {author} {\bibfnamefont
  {P.~S.}\ \bibnamefont {Julienne}}, \bibinfo {author} {\bibfnamefont {J.~E.}\
  \bibnamefont {Simsarian}}, \bibinfo {author} {\bibfnamefont {K.}~\bibnamefont
  {Helmerson}}, \bibinfo {author} {\bibfnamefont {S.~L.}\ \bibnamefont
  {Rolston}}, \ and\ \bibinfo {author} {\bibfnamefont {W.~D.}\ \bibnamefont
  {Phillips}},\ }\href {\doibase 10.1038/18395} {\bibfield  {journal} {\bibinfo
   {journal} {Nature}\ }\textbf {\bibinfo {volume} {398}},\ \bibinfo {pages}
  {218} (\bibinfo {year} {1999})}\BibitemShut {NoStop}%
\bibitem [{\citenamefont {Rolston}\ and\ \citenamefont
  {Phillips}(2002)}]{Rolston-NL-2002}%
  \BibitemOpen
  \bibfield  {author} {\bibinfo {author} {\bibfnamefont {S.~L.}\ \bibnamefont
  {Rolston}}\ and\ \bibinfo {author} {\bibfnamefont {W.~D.}\ \bibnamefont
  {Phillips}},\ }\href {\doibase 10.1038/416219a} {\bibfield  {journal}
  {\bibinfo  {journal} {Nature}\ }\textbf {\bibinfo {volume} {416}},\ \bibinfo
  {pages} {219} (\bibinfo {year} {2002})}\BibitemShut {NoStop}%
\bibitem [{\citenamefont {Pertot}\ \emph {et~al.}(2010)\citenamefont {Pertot},
  \citenamefont {Gadway},\ and\ \citenamefont {Schneble}}]{Pertot-10-PRL}%
  \BibitemOpen
  \bibfield  {author} {\bibinfo {author} {\bibfnamefont {D.}~\bibnamefont
  {Pertot}}, \bibinfo {author} {\bibfnamefont {B.}~\bibnamefont {Gadway}}, \
  and\ \bibinfo {author} {\bibfnamefont {D.}~\bibnamefont {Schneble}},\ }\href
  {\doibase 10.1103/PhysRevLett.104.200402} {\bibfield  {journal} {\bibinfo
  {journal} {Phys. Rev. Lett.}\ }\textbf {\bibinfo {volume} {104}},\ \bibinfo
  {pages} {200402} (\bibinfo {year} {2010})}\BibitemShut {NoStop}%
\bibitem [{\citenamefont {Campbell}\ \emph {et~al.}(2006)\citenamefont
  {Campbell}, \citenamefont {Mun}, \citenamefont {Boyd}, \citenamefont
  {Streed}, \citenamefont {Ketterle},\ and\ \citenamefont
  {Pritchard}}]{Gretchen-FWM}%
  \BibitemOpen
  \bibfield  {author} {\bibinfo {author} {\bibfnamefont {G.~K.}\ \bibnamefont
  {Campbell}}, \bibinfo {author} {\bibfnamefont {J.}~\bibnamefont {Mun}},
  \bibinfo {author} {\bibfnamefont {M.}~\bibnamefont {Boyd}}, \bibinfo {author}
  {\bibfnamefont {E.~W.}\ \bibnamefont {Streed}}, \bibinfo {author}
  {\bibfnamefont {W.}~\bibnamefont {Ketterle}}, \ and\ \bibinfo {author}
  {\bibfnamefont {D.~E.}\ \bibnamefont {Pritchard}},\ }\href {\doibase
  10.1103/PhysRevLett.96.020406} {\bibfield  {journal} {\bibinfo  {journal}
  {Phys. Rev. Lett.}\ }\textbf {\bibinfo {volume} {96}},\ \bibinfo {pages}
  {020406} (\bibinfo {year} {2006})}\BibitemShut {NoStop}%
\bibitem [{\citenamefont {Fallani}\ \emph {et~al.}(2004)\citenamefont
  {Fallani}, \citenamefont {De~Sarlo}, \citenamefont {Lye}, \citenamefont
  {Modugno}, \citenamefont {Saers}, \citenamefont {Fort},\ and\ \citenamefont
  {Inguscio}}]{Fallani-instable}%
  \BibitemOpen
  \bibfield  {author} {\bibinfo {author} {\bibfnamefont {L.}~\bibnamefont
  {Fallani}}, \bibinfo {author} {\bibfnamefont {L.}~\bibnamefont {De~Sarlo}},
  \bibinfo {author} {\bibfnamefont {J.~E.}\ \bibnamefont {Lye}}, \bibinfo
  {author} {\bibfnamefont {M.}~\bibnamefont {Modugno}}, \bibinfo {author}
  {\bibfnamefont {R.}~\bibnamefont {Saers}}, \bibinfo {author} {\bibfnamefont
  {C.}~\bibnamefont {Fort}}, \ and\ \bibinfo {author} {\bibfnamefont
  {M.}~\bibnamefont {Inguscio}},\ }\href {\doibase
  10.1103/PhysRevLett.93.140406} {\bibfield  {journal} {\bibinfo  {journal}
  {Phys. Rev. Lett.}\ }\textbf {\bibinfo {volume} {93}},\ \bibinfo {pages}
  {140406} (\bibinfo {year} {2004})}\BibitemShut {NoStop}%
\bibitem [{\citenamefont {Baharian}\ and\ \citenamefont {Baym}(2013)}]{GORDON}%
  \BibitemOpen
  \bibfield  {author} {\bibinfo {author} {\bibfnamefont {S.}~\bibnamefont
  {Baharian}}\ and\ \bibinfo {author} {\bibfnamefont {G.}~\bibnamefont
  {Baym}},\ }\href {\doibase 10.1103/PhysRevA.87.013619} {\bibfield  {journal}
  {\bibinfo  {journal} {Phys. Rev. A}\ }\textbf {\bibinfo {volume} {87}},\
  \bibinfo {pages} {013619} (\bibinfo {year} {2013})}\BibitemShut {NoStop}%
\bibitem [{\citenamefont {Koller}\ \emph {et~al.}(2016)\citenamefont {Koller},
  \citenamefont {Goldschmidt}, \citenamefont {Brown}, \citenamefont {Wyllie},
  \citenamefont {Wilson},\ and\ \citenamefont {Porto}}]{Koller-Looped}%
  \BibitemOpen
  \bibfield  {author} {\bibinfo {author} {\bibfnamefont {S.~B.}\ \bibnamefont
  {Koller}}, \bibinfo {author} {\bibfnamefont {E.~A.}\ \bibnamefont
  {Goldschmidt}}, \bibinfo {author} {\bibfnamefont {R.~C.}\ \bibnamefont
  {Brown}}, \bibinfo {author} {\bibfnamefont {R.}~\bibnamefont {Wyllie}},
  \bibinfo {author} {\bibfnamefont {R.~M.}\ \bibnamefont {Wilson}}, \ and\
  \bibinfo {author} {\bibfnamefont {J.~V.}\ \bibnamefont {Porto}},\ }\href
  {\doibase 10.1103/PhysRevA.94.063634} {\bibfield  {journal} {\bibinfo
  {journal} {Phys. Rev. A}\ }\textbf {\bibinfo {volume} {94}},\ \bibinfo
  {pages} {063634} (\bibinfo {year} {2016})}\BibitemShut {NoStop}%
\bibitem [{\citenamefont {Eckel}\ \emph {et~al.}(2014)\citenamefont {Eckel},
  \citenamefont {Lee}, \citenamefont {Jendrzejewski}, \citenamefont {Murray},
  \citenamefont {Clark}, \citenamefont {Lobb}, \citenamefont {Phillips},
  \citenamefont {Edwards},\ and\ \citenamefont
  {Campbell}}]{Eckel-Hysteresis-2014}%
  \BibitemOpen
  \bibfield  {author} {\bibinfo {author} {\bibfnamefont {S.}~\bibnamefont
  {Eckel}}, \bibinfo {author} {\bibfnamefont {J.~G.}\ \bibnamefont {Lee}},
  \bibinfo {author} {\bibfnamefont {F.}~\bibnamefont {Jendrzejewski}}, \bibinfo
  {author} {\bibfnamefont {N.}~\bibnamefont {Murray}}, \bibinfo {author}
  {\bibfnamefont {C.~W.}\ \bibnamefont {Clark}}, \bibinfo {author}
  {\bibfnamefont {C.~J.}\ \bibnamefont {Lobb}}, \bibinfo {author}
  {\bibfnamefont {W.~D.}\ \bibnamefont {Phillips}}, \bibinfo {author}
  {\bibfnamefont {M.}~\bibnamefont {Edwards}}, \ and\ \bibinfo {author}
  {\bibfnamefont {G.~K.}\ \bibnamefont {Campbell}},\ }\href {\doibase
  10.1038/nature12958} {\bibfield  {journal} {\bibinfo  {journal} {Nature}\
  }\textbf {\bibinfo {volume} {506}},\ \bibinfo {pages} {200} (\bibinfo {year}
  {2014})}\BibitemShut {NoStop}%
\bibitem [{\citenamefont {Ryu}\ \emph {et~al.}(2013)\citenamefont {Ryu},
  \citenamefont {Blackburn}, \citenamefont {Blinova},\ and\ \citenamefont
  {Boshier}}]{Ryu-SQUID}%
  \BibitemOpen
  \bibfield  {author} {\bibinfo {author} {\bibfnamefont {C.}~\bibnamefont
  {Ryu}}, \bibinfo {author} {\bibfnamefont {P.~W.}\ \bibnamefont {Blackburn}},
  \bibinfo {author} {\bibfnamefont {A.~A.}\ \bibnamefont {Blinova}}, \ and\
  \bibinfo {author} {\bibfnamefont {M.~G.}\ \bibnamefont {Boshier}},\ }\href
  {\doibase 10.1103/PhysRevLett.111.205301} {\bibfield  {journal} {\bibinfo
  {journal} {Phys. Rev. Lett.}\ }\textbf {\bibinfo {volume} {111}},\ \bibinfo
  {pages} {205301} (\bibinfo {year} {2013})}\BibitemShut {NoStop}%
\bibitem [{\citenamefont {Trenkwalder}\ \emph {et~al.}(2016)\citenamefont
  {Trenkwalder}, \citenamefont {Spagnolli}, \citenamefont {Semeghini},
  \citenamefont {Coop}, \citenamefont {Landini}, \citenamefont {Castilho},
  \citenamefont {Pezz\`{e}}, \citenamefont {Modugno}, \citenamefont {Inguscio},
  \citenamefont {Smerzi},\ and\ \citenamefont
  {Fattori}}]{Trenkwalder-ParSymmBreak-2016}%
  \BibitemOpen
  \bibfield  {author} {\bibinfo {author} {\bibfnamefont {A.}~\bibnamefont
  {Trenkwalder}}, \bibinfo {author} {\bibfnamefont {G.}~\bibnamefont
  {Spagnolli}}, \bibinfo {author} {\bibfnamefont {G.}~\bibnamefont
  {Semeghini}}, \bibinfo {author} {\bibfnamefont {S.}~\bibnamefont {Coop}},
  \bibinfo {author} {\bibfnamefont {M.}~\bibnamefont {Landini}}, \bibinfo
  {author} {\bibfnamefont {P.}~\bibnamefont {Castilho}}, \bibinfo {author}
  {\bibfnamefont {L.}~\bibnamefont {Pezz\`{e}}}, \bibinfo {author}
  {\bibfnamefont {G.}~\bibnamefont {Modugno}}, \bibinfo {author} {\bibfnamefont
  {M.}~\bibnamefont {Inguscio}}, \bibinfo {author} {\bibfnamefont
  {A.}~\bibnamefont {Smerzi}}, \ and\ \bibinfo {author} {\bibfnamefont
  {M.}~\bibnamefont {Fattori}},\ }\href {\doibase 10.1038/nphys3743} {\bibfield
   {journal} {\bibinfo  {journal} {Nat. Phys.}\ }\textbf {\bibinfo {volume}
  {12}},\ \bibinfo {pages} {826} (\bibinfo {year} {2016})}\BibitemShut
  {NoStop}%
\bibitem [{\citenamefont {Raghavan}\ \emph {et~al.}(2001)\citenamefont
  {Raghavan}, \citenamefont {Pu}, \citenamefont {Meystre},\ and\ \citenamefont
  {Bigelow}}]{RaghavanBigelow}%
  \BibitemOpen
  \bibfield  {author} {\bibinfo {author} {\bibfnamefont {S.}~\bibnamefont
  {Raghavan}}, \bibinfo {author} {\bibfnamefont {H.}~\bibnamefont {Pu}},
  \bibinfo {author} {\bibfnamefont {P.}~\bibnamefont {Meystre}}, \ and\
  \bibinfo {author} {\bibfnamefont {N.}~\bibnamefont {Bigelow}},\ }\href
  {\doibase 10.1016/S0030-4018(00)01163-9} {\bibfield  {journal} {\bibinfo
  {journal} {Opt. Commun.}\ }\textbf {\bibinfo {volume} {188}},\ \bibinfo
  {pages} {149 } (\bibinfo {year} {2001})}\BibitemShut {NoStop}%
\bibitem [{\citenamefont {Albiez}\ \emph {et~al.}(2005)\citenamefont {Albiez},
  \citenamefont {Gati}, \citenamefont {F\"olling}, \citenamefont {Hunsmann},
  \citenamefont {Cristiani},\ and\ \citenamefont
  {Oberthaler}}]{Albiez-Oberth-2005}%
  \BibitemOpen
  \bibfield  {author} {\bibinfo {author} {\bibfnamefont {M.}~\bibnamefont
  {Albiez}}, \bibinfo {author} {\bibfnamefont {R.}~\bibnamefont {Gati}},
  \bibinfo {author} {\bibfnamefont {J.}~\bibnamefont {F\"olling}}, \bibinfo
  {author} {\bibfnamefont {S.}~\bibnamefont {Hunsmann}}, \bibinfo {author}
  {\bibfnamefont {M.}~\bibnamefont {Cristiani}}, \ and\ \bibinfo {author}
  {\bibfnamefont {M.~K.}\ \bibnamefont {Oberthaler}},\ }\href {\doibase
  10.1103/PhysRevLett.95.010402} {\bibfield  {journal} {\bibinfo  {journal}
  {Phys. Rev. Lett.}\ }\textbf {\bibinfo {volume} {95}},\ \bibinfo {pages}
  {010402} (\bibinfo {year} {2005})}\BibitemShut {NoStop}%
\bibitem [{\citenamefont {Levy}\ \emph {et~al.}(2007)\citenamefont {Levy},
  \citenamefont {Lahoud}, \citenamefont {Shomroni},\ and\ \citenamefont
  {Steinhauer}}]{Levy-ACDC-2007}%
  \BibitemOpen
  \bibfield  {author} {\bibinfo {author} {\bibfnamefont {S.}~\bibnamefont
  {Levy}}, \bibinfo {author} {\bibfnamefont {E.}~\bibnamefont {Lahoud}},
  \bibinfo {author} {\bibfnamefont {I.}~\bibnamefont {Shomroni}}, \ and\
  \bibinfo {author} {\bibfnamefont {J.}~\bibnamefont {Steinhauer}},\ }\href
  {\doibase 10.1038/nature06186} {\bibfield  {journal} {\bibinfo  {journal}
  {Nature}\ }\textbf {\bibinfo {volume} {449}},\ \bibinfo {pages} {579}
  (\bibinfo {year} {2007})}\BibitemShut {NoStop}%
\bibitem [{\citenamefont {LeBlanc}\ \emph {et~al.}(2011)\citenamefont
  {LeBlanc}, \citenamefont {Bardon}, \citenamefont {McKeever}, \citenamefont
  {Extavour}, \citenamefont {Jervis}, \citenamefont {Thywissen}, \citenamefont
  {Piazza},\ and\ \citenamefont {Smerzi}}]{Leblanc-Joseph}%
  \BibitemOpen
  \bibfield  {author} {\bibinfo {author} {\bibfnamefont {L.~J.}\ \bibnamefont
  {LeBlanc}}, \bibinfo {author} {\bibfnamefont {A.~B.}\ \bibnamefont {Bardon}},
  \bibinfo {author} {\bibfnamefont {J.}~\bibnamefont {McKeever}}, \bibinfo
  {author} {\bibfnamefont {M.~H.~T.}\ \bibnamefont {Extavour}}, \bibinfo
  {author} {\bibfnamefont {D.}~\bibnamefont {Jervis}}, \bibinfo {author}
  {\bibfnamefont {J.~H.}\ \bibnamefont {Thywissen}}, \bibinfo {author}
  {\bibfnamefont {F.}~\bibnamefont {Piazza}}, \ and\ \bibinfo {author}
  {\bibfnamefont {A.}~\bibnamefont {Smerzi}},\ }\href {\doibase
  10.1103/PhysRevLett.106.025302} {\bibfield  {journal} {\bibinfo  {journal}
  {Phys. Rev. Lett.}\ }\textbf {\bibinfo {volume} {106}},\ \bibinfo {pages}
  {025302} (\bibinfo {year} {2011})}\BibitemShut {NoStop}%
\bibitem [{\citenamefont {Chang}\ \emph {et~al.}(2005)\citenamefont {Chang},
  \citenamefont {Qin}, \citenamefont {Zhang}, \citenamefont {You},\ and\
  \citenamefont {Chapman}}]{Chang-spinor-josephson}%
  \BibitemOpen
  \bibfield  {author} {\bibinfo {author} {\bibfnamefont {M.-S.}\ \bibnamefont
  {Chang}}, \bibinfo {author} {\bibfnamefont {Q.}~\bibnamefont {Qin}}, \bibinfo
  {author} {\bibfnamefont {W.}~\bibnamefont {Zhang}}, \bibinfo {author}
  {\bibfnamefont {L.}~\bibnamefont {You}}, \ and\ \bibinfo {author}
  {\bibfnamefont {M.~S.}\ \bibnamefont {Chapman}},\ }\href {\doibase
  10.1038/nphys153} {\bibfield  {journal} {\bibinfo  {journal} {Nat. Phys.}\
  }\textbf {\bibinfo {volume} {1}},\ \bibinfo {pages} {111} (\bibinfo {year}
  {2005})}\BibitemShut {NoStop}%
\bibitem [{\citenamefont {Tomkovi\ifmmode~\check{c}\else \v{c}\fi{}}\ \emph
  {et~al.}(2017)\citenamefont {Tomkovi\ifmmode~\check{c}\else \v{c}\fi{}},
  \citenamefont {Muessel}, \citenamefont {Strobel}, \citenamefont {L\"ock},
  \citenamefont {Schlagheck}, \citenamefont {Ketzmerick},\ and\ \citenamefont
  {Oberthaler}}]{Tomko-Oberth-2017}%
  \BibitemOpen
  \bibfield  {author} {\bibinfo {author} {\bibfnamefont {J.}~\bibnamefont
  {Tomkovi\ifmmode~\check{c}\else \v{c}\fi{}}}, \bibinfo {author}
  {\bibfnamefont {W.}~\bibnamefont {Muessel}}, \bibinfo {author} {\bibfnamefont
  {H.}~\bibnamefont {Strobel}}, \bibinfo {author} {\bibfnamefont
  {S.}~\bibnamefont {L\"ock}}, \bibinfo {author} {\bibfnamefont
  {P.}~\bibnamefont {Schlagheck}}, \bibinfo {author} {\bibfnamefont
  {R.}~\bibnamefont {Ketzmerick}}, \ and\ \bibinfo {author} {\bibfnamefont
  {M.~K.}\ \bibnamefont {Oberthaler}},\ }\href {\doibase
  10.1103/PhysRevA.95.011602} {\bibfield  {journal} {\bibinfo  {journal} {Phys.
  Rev. A}\ }\textbf {\bibinfo {volume} {95}},\ \bibinfo {pages} {011602}
  (\bibinfo {year} {2017})}\BibitemShut {NoStop}%
\bibitem [{\citenamefont {An}\ \emph {et~al.}(2017{\natexlab{c}})\citenamefont
  {An}, \citenamefont {Meier},\ and\ \citenamefont {Gadway}}]{An-Zigzag}%
  \BibitemOpen
  \bibfield  {author} {\bibinfo {author} {\bibfnamefont {F.~A.}\ \bibnamefont
  {An}}, \bibinfo {author} {\bibfnamefont {E.~J.}\ \bibnamefont {Meier}}, \
  and\ \bibinfo {author} {\bibfnamefont {B.}~\bibnamefont {Gadway}},\
  }\href@noop {} {\  (\bibinfo {year} {2017}{\natexlab{c}})},\ \Eprint
  {http://arxiv.org/abs/1705.09268} {arXiv:1705.09268} \BibitemShut {NoStop}%
\bibitem [{\citenamefont {Trombettoni}\ and\ \citenamefont
  {Smerzi}(2001)}]{Tromb-Breather}%
  \BibitemOpen
  \bibfield  {author} {\bibinfo {author} {\bibfnamefont {A.}~\bibnamefont
  {Trombettoni}}\ and\ \bibinfo {author} {\bibfnamefont {A.}~\bibnamefont
  {Smerzi}},\ }\href {\doibase 10.1103/PhysRevLett.86.2353} {\bibfield
  {journal} {\bibinfo  {journal} {Phys. Rev. Lett.}\ }\textbf {\bibinfo
  {volume} {86}},\ \bibinfo {pages} {2353} (\bibinfo {year}
  {2001})}\BibitemShut {NoStop}%
\bibitem [{\citenamefont {Anisimovas}\ \emph {et~al.}(2016)\citenamefont
  {Anisimovas}, \citenamefont {Ra\ifmmode \check{c}\else
  \v{c}\fi{}i\ifmmode~\bar{u}\else \={u}\fi{}nas}, \citenamefont {Str\"ater},
  \citenamefont {Eckardt}, \citenamefont {Spielman},\ and\ \citenamefont
  {Juzeli\ifmmode~\bar{u}\else \={u}\fi{}nas}}]{Ani-synthz-zigzag}%
  \BibitemOpen
  \bibfield  {author} {\bibinfo {author} {\bibfnamefont {E.}~\bibnamefont
  {Anisimovas}}, \bibinfo {author} {\bibfnamefont {M.}~\bibnamefont {Ra\ifmmode
  \check{c}\else \v{c}\fi{}i\ifmmode~\bar{u}\else \={u}\fi{}nas}}, \bibinfo
  {author} {\bibfnamefont {C.}~\bibnamefont {Str\"ater}}, \bibinfo {author}
  {\bibfnamefont {A.}~\bibnamefont {Eckardt}}, \bibinfo {author} {\bibfnamefont
  {I.~B.}\ \bibnamefont {Spielman}}, \ and\ \bibinfo {author} {\bibfnamefont
  {G.}~\bibnamefont {Juzeli\ifmmode~\bar{u}\else \={u}\fi{}nas}},\ }\href
  {\doibase 10.1103/PhysRevA.94.063632} {\bibfield  {journal} {\bibinfo
  {journal} {Phys. Rev. A}\ }\textbf {\bibinfo {volume} {94}},\ \bibinfo
  {pages} {063632} (\bibinfo {year} {2016})}\BibitemShut {NoStop}%
\bibitem [{\citenamefont {Tai}\ \emph {et~al.}(2017)\citenamefont {Tai},
  \citenamefont {Lukin}, \citenamefont {Rispoli}, \citenamefont {Schittko},
  \citenamefont {Menke}, \citenamefont {Borgnia}, \citenamefont {Preiss},
  \citenamefont {Grusdt}, \citenamefont {Kaufman},\ and\ \citenamefont
  {Greiner}}]{Tai-FluxLadderInt}%
  \BibitemOpen
  \bibfield  {author} {\bibinfo {author} {\bibfnamefont {M.~E.}\ \bibnamefont
  {Tai}}, \bibinfo {author} {\bibfnamefont {A.}~\bibnamefont {Lukin}}, \bibinfo
  {author} {\bibfnamefont {M.}~\bibnamefont {Rispoli}}, \bibinfo {author}
  {\bibfnamefont {R.}~\bibnamefont {Schittko}}, \bibinfo {author}
  {\bibfnamefont {T.}~\bibnamefont {Menke}}, \bibinfo {author} {\bibfnamefont
  {D.}~\bibnamefont {Borgnia}}, \bibinfo {author} {\bibfnamefont {P.~M.}\
  \bibnamefont {Preiss}}, \bibinfo {author} {\bibfnamefont {F.}~\bibnamefont
  {Grusdt}}, \bibinfo {author} {\bibfnamefont {A.~M.}\ \bibnamefont {Kaufman}},
  \ and\ \bibinfo {author} {\bibfnamefont {M.}~\bibnamefont {Greiner}},\ }\href
  {http://dx.doi.org/10.1038/nature22811} {\bibfield  {journal} {\bibinfo
  {journal} {Nature}\ }\textbf {\bibinfo {volume} {546}},\ \bibinfo {pages}
  {519} (\bibinfo {year} {2017})}\BibitemShut {NoStop}%
\bibitem [{\citenamefont {Khomeriki}\ and\ \citenamefont
  {Flach}(2016)}]{DiamondLattice-2016}%
  \BibitemOpen
  \bibfield  {author} {\bibinfo {author} {\bibfnamefont {R.}~\bibnamefont
  {Khomeriki}}\ and\ \bibinfo {author} {\bibfnamefont {S.}~\bibnamefont
  {Flach}},\ }\href {\doibase 10.1103/PhysRevLett.116.245301} {\bibfield
  {journal} {\bibinfo  {journal} {Phys. Rev. Lett.}\ }\textbf {\bibinfo
  {volume} {116}},\ \bibinfo {pages} {245301} (\bibinfo {year}
  {2016})}\BibitemShut {NoStop}%
\bibitem [{\citenamefont {Lin}\ \emph {et~al.}(2014)\citenamefont {Lin},
  \citenamefont {Zhang},\ and\ \citenamefont {Scarola}}]{Vito-Flatband-14}%
  \BibitemOpen
  \bibfield  {author} {\bibinfo {author} {\bibfnamefont {F.}~\bibnamefont
  {Lin}}, \bibinfo {author} {\bibfnamefont {C.}~\bibnamefont {Zhang}}, \ and\
  \bibinfo {author} {\bibfnamefont {V.~W.}\ \bibnamefont {Scarola}},\ }\href
  {\doibase 10.1103/PhysRevLett.112.110404} {\bibfield  {journal} {\bibinfo
  {journal} {Phys. Rev. Lett.}\ }\textbf {\bibinfo {volume} {112}},\ \bibinfo
  {pages} {110404} (\bibinfo {year} {2014})}\BibitemShut {NoStop}%
\bibitem [{\citenamefont {Huber}\ and\ \citenamefont
  {Altman}(2010)}]{Ehud-FlatBand-10}%
  \BibitemOpen
  \bibfield  {author} {\bibinfo {author} {\bibfnamefont {S.~D.}\ \bibnamefont
  {Huber}}\ and\ \bibinfo {author} {\bibfnamefont {E.}~\bibnamefont {Altman}},\
  }\href {\doibase 10.1103/PhysRevB.82.184502} {\bibfield  {journal} {\bibinfo
  {journal} {Phys. Rev. B}\ }\textbf {\bibinfo {volume} {82}},\ \bibinfo
  {pages} {184502} (\bibinfo {year} {2010})}\BibitemShut {NoStop}%
\bibitem [{\citenamefont {Leykam}\ \emph {et~al.}(2013)\citenamefont {Leykam},
  \citenamefont {Flach}, \citenamefont {Bahat-Treidel},\ and\ \citenamefont
  {Desyatnikov}}]{FlatBand-Nonlinear}%
  \BibitemOpen
  \bibfield  {author} {\bibinfo {author} {\bibfnamefont {D.}~\bibnamefont
  {Leykam}}, \bibinfo {author} {\bibfnamefont {S.}~\bibnamefont {Flach}},
  \bibinfo {author} {\bibfnamefont {O.}~\bibnamefont {Bahat-Treidel}}, \ and\
  \bibinfo {author} {\bibfnamefont {A.~S.}\ \bibnamefont {Desyatnikov}},\
  }\href {\doibase 10.1103/PhysRevB.88.224203} {\bibfield  {journal} {\bibinfo
  {journal} {Phys. Rev. B}\ }\textbf {\bibinfo {volume} {88}},\ \bibinfo
  {pages} {224203} (\bibinfo {year} {2013})}\BibitemShut {NoStop}%
\bibitem [{\citenamefont {Aleiner}\ \emph {et~al.}(2010)\citenamefont
  {Aleiner}, \citenamefont {Altshuler},\ and\ \citenamefont
  {Shlyapnikov}}]{aleiner:finite_temperature_disorder_2010}%
  \BibitemOpen
  \bibfield  {author} {\bibinfo {author} {\bibfnamefont {I.~L.}\ \bibnamefont
  {Aleiner}}, \bibinfo {author} {\bibfnamefont {B.~L.}\ \bibnamefont
  {Altshuler}}, \ and\ \bibinfo {author} {\bibfnamefont {G.~V.}\ \bibnamefont
  {Shlyapnikov}},\ }\href {\doibase 10.1038/nphys1758} {\bibfield  {journal}
  {\bibinfo  {journal} {Nat. Phys.}\ }\textbf {\bibinfo {volume} {6}},\
  \bibinfo {pages} {900} (\bibinfo {year} {2010})}\BibitemShut {NoStop}%
\bibitem [{\citenamefont {Deissler}\ \emph {et~al.}(2010)\citenamefont
  {Deissler}, \citenamefont {Zaccanti}, \citenamefont {Roati}, \citenamefont
  {D'Errico}, \citenamefont {Fattori}, \citenamefont {Modugno}, \citenamefont
  {Modugno},\ and\ \citenamefont
  {Inguscio}}]{Deissler-DisorderWithInteractions-2010}%
  \BibitemOpen
  \bibfield  {author} {\bibinfo {author} {\bibfnamefont {B.}~\bibnamefont
  {Deissler}}, \bibinfo {author} {\bibfnamefont {M.}~\bibnamefont {Zaccanti}},
  \bibinfo {author} {\bibfnamefont {G.}~\bibnamefont {Roati}}, \bibinfo
  {author} {\bibfnamefont {C.}~\bibnamefont {D'Errico}}, \bibinfo {author}
  {\bibfnamefont {M.}~\bibnamefont {Fattori}}, \bibinfo {author} {\bibfnamefont
  {M.}~\bibnamefont {Modugno}}, \bibinfo {author} {\bibfnamefont
  {G.}~\bibnamefont {Modugno}}, \ and\ \bibinfo {author} {\bibfnamefont
  {M.}~\bibnamefont {Inguscio}},\ }\href {\doibase 10.1038/nphys1635}
  {\bibfield  {journal} {\bibinfo  {journal} {Nat. Phys.}\ }\textbf {\bibinfo
  {volume} {6}},\ \bibinfo {pages} {354} (\bibinfo {year} {2010})}\BibitemShut
  {NoStop}%
\end{thebibliography}


\end{document}

\title{Supplementary materials for ``Correlated dynamics in a synthetic lattice of momentum states''}
\author{Fangzhao~Alex~An}
\thanks{These authors contributed equally to this work}
\author{Eric~J.~Meier}
\thanks{These authors contributed equally to this work}
\author{Jackson~Ang'ong'a}
\author{Bryce~Gadway}
\email{bgadway@illinois.edu}
\affiliation{Department of Physics, University of Illinois at Urbana-Champaign, Urbana, IL 61801-3080, USA}
\date{\today}

\renewcommand\thefigure{S\arabic{figure}}
\maketitle

\section{Fitting and simulation procedure}

We determine the tunneling energy $t$, average mean-field shift $U$, and initial condensate momentum $p_i$ by performing a combined fit of all experimental data in Figs.~2(a-d) of the main text to numerical simulations. Specifically, we minimize the combined residuals of the two Bragg spectroscopy data sets for the $0\rightarrow\pm 1$ transitions and the two (positive and negative) sweep data sets as a function of the simulation parameters $t$, $U$, and $p_i$.

For each data set, we generated simulation curves by solving the nonlinear Schr\"{o}dinger equation of Eq.~2 with appropriate detuning/well imbalance $\Delta$ (fixed $\Delta$ for the Bragg data and swept $\Delta$ for the sweep data). The simulations assume a given tunneling energy $t$, initial condensate momentum $p_i$ (relating to an added intersite bias), and peak mean-field energy $U_0$. In the spirit of a local density approximation, we take into account the inhomogeneous density and inhomogeneous mean-field energy of our trapped atomic condensates by assuming a Thomas-Fermi density distribution and taking a population-weighted sum of simulation curves based on a homogeneous mean-field energy $U$, ranging from $0$ to $U_0$.

We took the residual sum of squares for each data set, and weighted each of these by the number of data points in the respective data set to give equal weighting to all data sets before summing up all of the values to get one fitting metric.
We then minimized this weighted sum with respect to the input values $t$, $U_0$, and $p_i$, obtaining tunneling energy $t/\hbar \approx 2\pi\times 390$~Hz, average mean-field energy $U/\hbar \approx 2\pi\times 1810$~Hz relating to an inhomogeneous distribution with peak mean-field energy $U_0/\hbar \approx 2\pi\times 3170$~Hz, and an initial condensate momentum $p_i \approx -0.018 \, \hbar k$.

\section{Calculation of adiabatic energy levels}

Even ignoring quantum fluctuations and nonclassical correlations, the nonlinear momentum-space interactions can result in interesting behavior of the two-mode condensate field, as described in the main text. Here, we briefly describe the calculations (following the work of Ref.~\cite{GORDON}) used to determine the adiabatic energy levels and their model projections, as shown in Fig.~2 of the main text. The identification of cusps, or swallow-tails, in the adiabatic energy level structure provides insight into the mechanism underlying the observed dynamical phenomenon of nonlinear self-trapping.

We approximate the state of our two-mode Bose--Einstein condensate (with possible single particle states represented in terms of superpositions of the left ($|L\rangle \equiv |p/2\hbar k = 0\rangle$) and right well ($|R\rangle \equiv |p/2\hbar k = 1\rangle$) orbitals) in terms of the field $|\psi (\tau) \rangle = \sqrt{N} [\phi_L (\tau) |L\rangle +\phi_R (\tau) |R\rangle]$, where $\tau$ is the time variable, and the complex amplitudes $\phi_L$ and $\phi_R$ relate to populations $N_L = N|\phi_L|^2$ and $N_R = N|\phi_R|^2$ in the left and right well. As described in the main text, the nonlinear Schr\"{o}dinger equation describing the evolution of the condensate wave function results in the two coupled differential equations
\begin{align}
i\hbar \dot{\phi}_L &= U [2 - |\phi_L|^2]\phi_L - t \phi_R \\
i\hbar \dot{\phi}_R &= U [2 - |\phi_L|^2]\phi_R - t \phi_L + \Delta (\tau)\phi_R \ ,
\end{align}
where $U$ is the homogeneous mean-field energy relating to direct interactions, $t$ is the real-valued tunneling energy, and $\Delta$ is the site-to-site energy bias. In experiment, the bias is swept in time as $\Delta (\tau) = \Delta_i - 2\Delta_i (\tau/T)$ for a sweep duration $T = 1$~ms.

Intuition into the dynamical response of this system is gained by first considering the steady-state solutions to these coupled equations for various conditions of $U$, $t$, and $\Delta$. We determine the adiabatic energy levels relating to steady state solutions of the form $|\psi (\tau) \rangle = e^{-i\mu\tau} |\psi (0) \rangle$ by solving
\begin{align}
\mu \phi_L &= [U(2 - |\phi_L|^2)]\phi_L - t \phi_R
\label{eee} \\
\mu \phi_R &= [\Delta + U(2 - |\phi_R|^2)]\phi_R - t \phi_L \ ,
\label{fff}
\end{align}
where $\mu$ is the chemical potential.

Solution of these coupled equations determines the adiabatic eigenstate solutions, their energy values, and their projections onto the two available momentum modes (wells). For a given combination of $U$, $t$, and $\Delta$ values, between two and four solutions with distinct $\mu$ and $|\phi_L|$ combinations (and with $|\phi_R|^2 = 1 - |\phi_L|^2$ fixing $|\phi_R|$) can be found, with the presence of more than two solutions relating to cusp or swallow-tail structures in the energy band structures. The projected probabilities onto the left and right well states $|L\rangle$ and $|R\rangle$ are given by
\begin{equation}
|\phi_L|^2 = 1 - |\phi_R|^2 = \frac{\Delta + U - \mu}{\Delta + 2 (U-\mu)}  \ .
\end{equation}

The eigenstate solutions, having different $\mu$ values in general, are expressed in terms of these projected probabilities as $|\mathrm{I}\rangle = \sqrt{N} [ |\phi_L| |L\rangle - |\phi_R| |R\rangle ]$ (the ground state) and $|\mathrm{II}\rangle = \sqrt{N} [ |\phi_L| |L\rangle + |\phi_R| |R\rangle ]$ (the excited state). The relevant $\mu$ values may be determined by solving the determinant
\begin{equation}
\Bigg | \left(
   \begin{array}{cc}
     U(2-|\phi_L|^2) - \mu & -t \\
     -t & \Delta + U(2-|\phi_R|^2) - \mu \\
   \end{array}
 \right) \Bigg| = 0
\end{equation}
in combination with solution of Eq.~\ref{eee} and Eq.~\ref{fff}. We note that in the main text (Fig.~2), these $\mu$ values are labeled as the energies $E$ of the adiabatic energy levels.
%Finally, the adiabatic energy levels as plotted in the main text can be related to the various chemical potentials as
%\begin{equation}
%E = \mu - \frac{1}{2}U (1 + 2 |\phi_0|^2 |\phi_1|^2)  \ .
%\end{equation}

\section{Effect of interactions on quantum correlations in the momentum-space double well}

The role of quantum fluctuations of the momentum-space double well site populations has been ignored in the main text, yet significant quantum correlations can develop in such a system as a result of interactions. In particular, the momentum-space interaction shown to lead to the onset of nonlinear self-trapping in the main text can also be harnessed to generate squeezing of the momentum-space distribution. Specifically, a generic two-mode quantum system featuring mode-dependent interactions can be mapped onto an effective collective spin Hamiltonian featuring a one-axis twisting term that can generate squeezed collective spin states~\cite{RaghavanBigelow,Esteve-Squeezing,Jo-squeezing,Gross-Squeezing}.

We consider a system of particles distributed between two momentum modes (labeled mode $0$ and mode $1$) with a fixed total particle number $N = N_0 + N_1$. In the main text, when simulating the experimental data for the case of $N \sim 10^5$ atoms and only two sites, we ignored quantum fluctuations of the atomic populations. This assumed that for a given site $n$ with particle number $N_n$, $\langle \hat{c}_n \rangle \sim \langle \hat{c}_n^\dagger \rangle \sim \sqrt{N_n}$, where $\hat{c}_n$ ($\hat{c}^\dagger_n$) is the annihilation (creation) operator for the mode $n$. This assumption allowed us to describe the system by a greatly simplified multimode Gross-Pitaevskii equation~\cite{Trippenbach-FWM-theory}. More generally, following the treatment of Ref.~\cite{RaghavanBigelow}, we may include the effect of quantum fluctuations by considering instead a second quantized Hamiltonian $H = H_\text{sp} + H_\text{int}$ that describes two-body interactions of atoms in the same and different modes ($H_\text{int}$), as well as the engineered synthetic lattice tunneling and site energies ($H_\text{sp}$). For the double well, these contributions may be written as
\begin{equation}
H_\text{sp} = \varepsilon_0 \hat{c}^\dagger_0 \hat{c}_0 + \varepsilon_1 \hat{c}^\dagger_1 \hat{c}_1 - t (\hat{c}^\dagger_0 \hat{c}_1 + \hat{c}^\dagger_1 \hat{c}_0)
\end{equation}
and
\begin{equation}
H_\text{int} = \frac{u_{00}}{2}(\hat{c}^\dagger_0 \hat{c}^\dagger_0 \hat{c}_0 \hat{c}_0) + \frac{u_{11}}{2}(\hat{c}^\dagger_1 \hat{c}^\dagger_1 \hat{c}_1 \hat{c}_1) + u_{01}(\hat{c}^\dagger_0 \hat{c}_0 \hat{c}^\dagger_1  \hat{c}_1) \ .
\end{equation}
Here, $t$ is the tunneling energy, which we assume to be real, and $\varepsilon_0$ and $\varepsilon_1$ are the site energies (with energy bias $\Delta = \varepsilon_1 - \varepsilon_0$), which may be controlled through the Bragg frequency detuning. The interaction coefficients are given by $u_{00} = u_{11} = U/N$ (for the assumed uniform mean-field energy $U$ and total particle number $N$) and $u_{01} = 2U/N$, with the difference arising from the added exchange interactions~\cite{Trippenbach-FWM-theory}.

\begin{figure}[t!]
\includegraphics[width=\columnwidth]{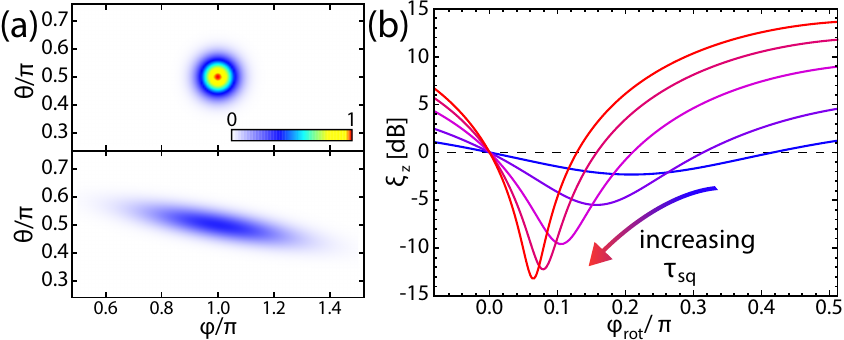}
\caption{\label{FIG:figs1}
\textbf{Squeezing in a momentum-space double well.}
(\textbf{a})~Visualization of many-particle ($N = 100$) spin states $|\Psi\rangle$ through their overlap with different coherent spin states $|\langle \theta,\varphi|\Psi\rangle|^2$. Shown are the cases of an initial coherent spin state $|\pi/2,\pi\rangle$ (upper plot), and the transformed state after evolution under $H_{\mathrm{sq}}$ for a time $\kappa \tau_{\mathrm{sq}}/\hbar = 0.0173 \pi$ (lower plot).
(\textbf{b})~Squeezing along the $\hat{z}$-axis, $\xi_z$, for different evolution times $\kappa \tau_{\mathrm{sq}}/\hbar = \{0.1,0.25,0.5,0.75,1\}\times 0.0173 \pi$ (solid lines, with colors varying from blue to red) and for different angles of rotation $\varphi_{\mathrm{rot}}$ of the final distribution about $\hat{J}_x$.
}
\end{figure}

Analysis of this interacting two-mode system is simplified by considering the atomic distribution in terms of a collective spin. Each atom is imbued with an effective spin-$1/2$ degree of freedom relating to the two momentum modes that may be occupied. The total state of the system may be considered as an effectively large, collective spin with maximum length $N/2$. One may define effective angular momentum operators relating to the coherences and macroscopic occupations $N_0$ and $N_1$ of two possible momentum orders (ignoring thermal and quantum depletion),
given by
\begin{equation}
\hat{J}_x = (\hat{c}_0^\dag \hat{c}_1 + \hat{c}_1^\dag \hat{c}_0)/2
\end{equation}
\begin{equation}
\hat{J}_y = i(\hat{c}_0^\dag \hat{c}_1 - \hat{c}_1^\dag \hat{c}_0)/2
\end{equation}
\begin{equation}
\hat{J}_z = (\hat{c}_1^\dag \hat{c}_1 - \hat{c}_0^\dag \hat{c}_0)/2 \ .
\end{equation}
In this modified description in terms of a collective spin, we will consider a basis of Dicke states $|j,m\rangle$ having total spin $j = N/2$ and $\hat{z}$-projection $m = (N_1 - N_0) / 2$, describing the collective two-mode number states. Uncorrelated coherent spin states (CSSs) of the form
\begin{equation}
|\theta , \varphi\rangle = \sum_{m = -j}^{j} f_{m}^j (\theta) e^{-i(j+m)\varphi}|j,m\rangle \ ,
\end{equation}
for $f_{m}^j (\theta) = \binom{2j}{j+m}^{1/2}\cos(\theta/2)^{j-m}\sin(\theta/2)^{j+m}$, result from a global rotation of spin-polarized states $|j,j\rangle$ about the spin vector $\hat{n}_\varphi = \cos(\varphi) \hat{J}_x + \sin(\varphi) \hat{J}_y$ by an angle $\theta$~\cite{CSS-RMP}. In this description in terms of a collective spin, the momentum-space interaction gives rise to an effective nonlinear squeezing Hamiltonian $H_{\mathrm{sq}} = \kappa \hat{J}_z^2$ (after removing the contribution from mode-independent interactions), where $\kappa = (u_{00} + u_{11})/2 - u_{01} = -U/N$~\cite{RaghavanBigelow}.

When combined with single particle manipulations of the effective spin degree of freedom through control of $H_{sp}$, the interactions can be used to generate correlations and entanglement in the double well system. We examine the case of how interactions modify an initially prepared CSS $|\pi/2,\pi\rangle$ aligned along $-\hat{J}_x$. For short evolution times, the nonlinear Hamiltonian $H_{\mathrm{sq}}$ leads to a ``shearing'' of such coherent states. This is depicted in Fig.~\pref{FIG:figs1}{(a)}, for the initial CSS (upper plot) and the sheared, non-classical squeezed state after a time $\kappa\tau_{\mathrm{sq}}/\hbar = 0.0173 \pi$ (lower plot), through the overlap of these states with CSSs of varying $\theta$ and $\varphi$ values. For ease of calculation, dynamics are shown for the case of only $N = 100$ atoms ($j = 50$).

Figure~\pref{FIG:figs1}{(b)} shows, for sheared distributions relating to various
evolution times $\tau_{\mathrm{sq}}$, the $\hat{z}$-axis squeezing parameter $\xi_z = 2 j \frac{\langle \Delta\hat{J}_z^2\rangle}{j^2 - \langle\hat{J}_z\rangle^2}$ as a function of rotation angle $\varphi_\text{rot}$ about the $\hat{J}_x$ spin axis. For typical experimental parameter values ($N = 10^5$ atoms, $U / \hbar = 2 \pi \times 1.5$~kHz), an optimal squeezing of $\xi_z^\text{min} \approx (3j^2)^{-1/3}$, relating to $-33$~dB (after appropriate rotation of the sheared distribution about the $\hat{J}_x$ spin axis), would be expected after a total duration $\tau_{\mathrm{sq}} \approx 1.13 j^{1/3}(\hbar/U) \approx 6.6$~ms~\cite{RaghavanBigelow}. The same squeezing could be achieved in a shorter time through a temporary modification of the atomic scattering length, such as through a magnetically tunable Feshbach resonance. Shorter squeezing timescales would help to mitigate effects owing to spatial separation of the momentum orders, which would in practice degrade and limit the available squeezing. Additionally, refocusing $\pi$ (echo) pulses or twist-and-turn squeezing schemes~\cite{Muessel-TWIST} can be used to maintain spatial overlap of the momentum states.

%\bibliographystyle{apsrev4-1}
%\bibliography{BibsIntSupp}
